\documentclass[journal=jacsat,manuscript=article]{achemso}

\usepackage{amsmath, amssymb, amsthm, bm}

\usepackage{lineno}
\usepackage{comment}
\usepackage{xcolor}

\usepackage[T1]{fontenc}         
\usepackage{standalone}
\usepackage[utf8]{inputenc} 
\usepackage{color}
\usepackage{graphicx}
\usepackage{amsmath, amssymb, amsthm, bm}
\usepackage{float}
\usepackage{lineno}
\usepackage{chemformula}
\usepackage{xymtex}
\usepackage{dcolumn}
\usepackage{multirow}
\usepackage[version=4]{mhchem}
\usepackage{booktabs}  
\usepackage{subcaption}
\usepackage{threeparttable}
\usepackage{xr}
\author{Haidi Wang}
\affiliation{School of Physics, Hefei University of Technology, Hefei, Anhui 230009, China}
\author{Yufan Yao}
\affiliation{Department of Chemical Physics, and Hefei National Research Center for Physical Sciences at the Microscale, University of Science and Technology of China, Hefei, Anhui 230026,  China}

\author{Haonan Song}
\affiliation{School of Physics, Hefei University of Technology, Hefei, Anhui 230009, China}

\author{Huimiao Wang}
\affiliation{School of Rare Earths, University of Science and Technology of China, Hefei, Anhui 230026}

\author{Xiaofeng Liu}
\affiliation{School of Physics, Hefei University of Technology, Hefei, Anhui 230009, China}

\author{Zhao Chen}
\affiliation{School of Physics, Hefei University of Technology, Hefei, Anhui 230009, China}

\author{Weiwei Chen}
\affiliation{School of Physics, Hefei University of Technology, Hefei, Anhui 230009, China}

\author{Weiduo Zhu}
\email{weiduozhu@hfut.edu.cn}
\affiliation{School of Physics, Hefei University of Technology, Hefei, Anhui 230009, China}

\author{Zhongjun Li}
\email{zjli@hfut.edu.cn}
\affiliation{School of Physics, Hefei University of Technology, Hefei, Anhui 230009, China}

\author{Jinlong Yang}
\email{jlyang@ustc.edu.cn}
\affiliation{Department of Chemical Physics, and Hefei National Research Center for Physical Sciences at the Microscale, University of Science and Technology of China, Hefei, Anhui 230026,  China}

\title[]
{Uni2D: A Universal Machine Learning Interatomic Potential for Two-Dimensional Materials}

\abbreviations{2D, ML, PES, DFT}

\begin{document}
	
	
	\begin{tocentry}
		
	\includegraphics[width=1\textwidth]{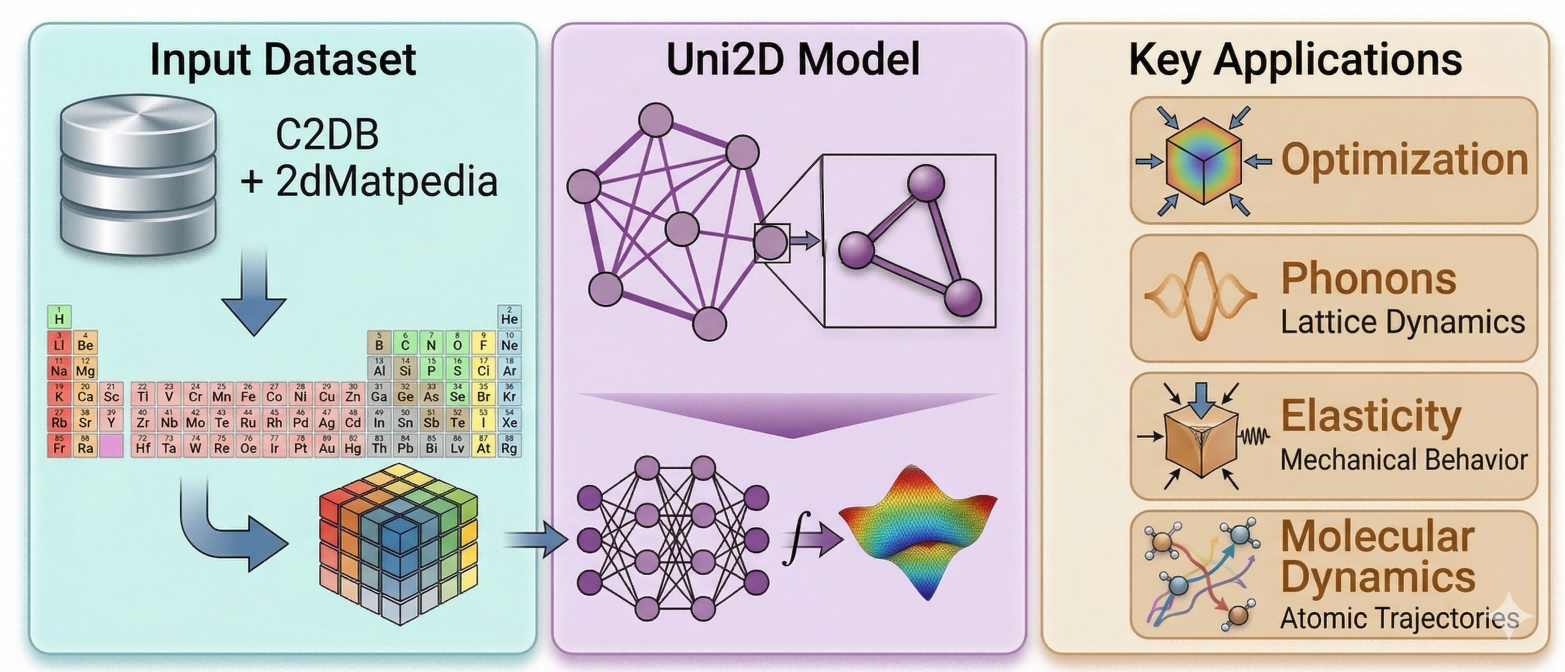} 
		
	\end{tocentry}
	
	
\begin{abstract}
Accurate interatomic potentials (IAPs) are essential for modeling the potential energy surfaces (PES) that govern atomic interactions in materials. However, most existing IAPs are developed for bulk materials and often struggle to accurately and efficiently capture the diverse chemical environments of two-dimensional (2D) materials, which limits large-scale simulation and design of emerging 2D systems. To address this challenge, we develop Uni2D, an interatomic potential tailored for 2D materials. The Uni2D model is trained on a dataset comprising approximately 327,000 structure–energy–force–stress mappings derived from about 20,000 distinct 2D materials, covering 89 chemical elements. The model demonstrates reliable predictive performance for energies, forces, and stresses, and demonstrates quantitatively robust accuracy in tasks such as structural relaxation, equation-of-state calculations, and molecular dynamics simulations, making the model suitable for high-throughput screening of 2D materials. For derived properties, including elastic properties, lattice dynamics, and other screening-related metrics, the model provides qualitative to semi-quantitative predictions that remain useful for trend analysis and preliminary evaluation. To enhance usability, we further introduce an intelligent agent powered by a large language model (LLM), enabling automated workflows and natural language interaction for 2D materials simulations. Our work provides an efficient and accessible framework for high-throughput screening and computational exploration of 2D materials.
\end{abstract}
	
	\section{Introduction}
	In recent years, atomic simulations have become an essential tool in materials science and condensed matter physics for discovering and designing novel material systems. In molecular dynamics simulations, IAPs\cite{BECKER2013277} are core tools used to approximate atomic interactions and construct PESs, determining the ability to accurately describe the energies and interactions of atomic systems. These PESs are crucial for a wide range of physical and chemical phenomena in materials, including phase transitions\cite{Wales2006Landscapes,PhysRevB.95.094203}, thermodynamic properties\cite{PhysRevE.63.041201}, crystal structure prediction\cite{Wang2022CSP,Wang2020DeepPotential,10.1063/5.0100505}, and so on. Thus, the accuracy of IAPs directly impacts the reliability and scientific value of these simulations.
	
	Traditional IAP models rely on a limited number of parameters to describe atomic interactions, such as the Lennard-Jones potential, embedded atom model\cite{PhysRevB.33.7983} (EAM), and Tersoff potential\cite{PhysRevB.37.6991}. These classical models have been successful in certain specific material systems, but their applicability is often limited to particular chemical compositions and physical conditions, lacking the universality required for complex chemical environments. As an alternative, \textit{Ab initio} molecular dynamics\cite{10.1073/pnas.0500193102} (AIMD) can accurately describe interatomic interactions, but its computational cost is prohibitively high for large-scale, long-duration simulations. To overcome these limitations and strike a balance between accuracy and computational efficiency, machine learning-based interatomic potentials\cite{10.1063/1.3553717, PhysRevLett.98.146401, THOMPSON2015316, PhysRevLett.120.143001,PhysRevB.95.094203} (ML-IAPs) have been rapidly developed in recent years, providing more efficient and accurate solutions for materials simulations.
	
	Over the past decade, significant advancements have been made in ML-IAPs. Early methods primarily relied on handcrafted features, with notable examples including the Behler-Parrinello symmetry functions and Gaussian approximation potentials (GAP)\cite{PhysRevLett.104.136403}. These approaches defined symmetry descriptors of the local atomic environment (such as bond lengths, bond angles, etc.) manually and used them as inputs to shallow neural networks or kernel methods to learn the potential energy surface from density functional theory (DFT) data. However, these methods were limited by feature engineering, which restricted their generalization capabilities. With the development of deep learning techniques, graph neural networks (GNNs) have gradually become the dominant framework. In 2017, SchNet\cite{10.1063/1.5019779} pioneered the use of continuous filter convolutions to model atomic systems, marking the beginning of deep learning-based IAP methods. In 2020, DimeNet\cite{NEURIPS2021_82489c97} introduced a scheme that explicitly captured three-body interactions through directional message passing, further enhancing the model’s ability to capture complex interactions. In 2021, NequIP\cite{Batzner2022E3Equivariant} introduced equivariant neural networks (E(3)-GNN), which improved both the physical consistency and predictive accuracy of the model by strictly enforcing physical symmetries. In recent years, ML-IAP models have continued to evolve in terms of higher-order interactions and generalizability. For instance, MACE (2022)\cite{NEURIPS2022_4a36c3c5} employed a multi-body attention mechanism to model four-body and higher-order interactions, significantly broadening the applicability of the model. Models such as CHGNet\cite{deng_2023_chgnet}, MatterSim\cite{yang2024mattersim}, and eqV2\cite{eqv22024} have leveraged large-scale pretraining to achieve transfer learning across elemental systems, further enhancing the adaptability and predictive power of the models. These advancements signify a shift toward more accurate, general, and efficient ML-IAPs, providing new tools and approaches for materials simulation and discovery \cite{loew2024uiap, https://doi.org/10.1002/mgea.58, 10.1063/5.0199743, doi:10.1021/acsami.4c03815,ju2025application}.
	
	Despite the rapid advancements in ML-IAPs that have significantly enhanced the capabilities of materials simulations, frameworks like M3GNet, CHGNet, MatterSim, and other models in the Matbench platform\cite{riebesell2024matbenchdiscoveryframework,matbench_discovery} have demonstrated outstanding performance in modeling bulk materials, their applicability and generalization to 2D materials remain under-explored. Due to their unique physicochemical properties—including significant surface effects, weaker interlayer coupling, and bonding characteristics distinct from three-dimensional materials—existing ML-IAP models may struggle to accurately capture the behavior of 2D systems. Consequently, directly applying models developed for three-dimensional materials to 2D systems presents substantial challenges. Additionally, existing generalized ML-IAP models are typically trained on datasets focused on bulk materials, lacking sufficient diversity to capture the atomic chemical environments of 2D materials, which limits their accuracy and generality in these contexts.
	
	To address these challenges, this study aims to develop a universal ML-IAP model for 2D materials named Uni2D within the MatterSim framework. We leveraged a diverse dataset specifically tailored to 2D materials to construct a model capable of generalizing across various 2D systems, providing high-accuracy predictions of their key physical properties and the design of new materials. In addition, the integration of this model with a large language model \cite{chiang2024llamplargelanguagemodel, jia2024llmatdesignautonomousmaterialsdiscovery} (LLM)-based agent interface not only advances the application of the Uni2D model to 2D materials but also paves the way for high-throughput screening and rational design of 2D materials systems, promising significant impacts on materials science research and engineering applications.
	\section{Methodology}
	
	\subsection{Model Architecture}
	 In this study, we adopt the MatterSim framework, which builds upon the M3GNet architecture\cite{m3gnet}, to train the Uni2D model. A key feature of M3GNet is the incorporation of three-body interactions within the message-passing framework, enabling more accurate updates of atomic and bond features by explicitly accounting for angular-dependent interactions. The model represents materials as graphs while preserving essential physical constraints and symmetries, including energy and force continuity, as well as invariance of energy under translation, rotation, and permutation of atomic configurations\cite{KHORSHIDI2016310}. Additional architectural details are provided in the Supporting Information (SI).
	
	\subsection{Dataset Preparation}
	To develop a robust and widely applicable ML-IAP model, comprehensive sampling of the PES of 2D materials is essential. Currently, several 2D materials databases, such as C2DB\cite{Gjerding_2021}, 2dMatpedia\cite{Zhou2019_2DMatPedia}, MC2D\cite{Campi2023_MaterialsCloud2D}, and V2DB\cite{Sorkun2020}, have amassed a substantial number of material structures. Among these, the C2DB database and the 2dMatpedia database contain approximately 15,000 and 4,000 structural entries, respectively. These databases encompass a wide range of prototypical 2D materials, including graphene\cite{doi:10.1126/science.1102896}, transition metal dichalcogenides (TMDs)\cite{Manzeli2017TMDs}, hexagonal boron nitride (h-BN)\cite{Zhang2017hBN}, phosphorene\cite{Liu2014b}, and various other layered compounds. However, the relaxed structures in these databases, obtained through first-principles calculations based on experimental or theoretical configurations, predominantly correspond to equilibrium structures that lie close to local energy minima and often exhibit relatively high symmetry. As a result, the sampled configurations mainly represent near-equilibrium states of the PES, which may limit the coverage of off-equilibrium configurations required for training robust ML-IAP models. To broaden the configuration sampling, we perform data augmentation on the existing structures of the C2DB and 2dMatpedia databases to expand the accessible configuration space. The detailed procedures for lattice strain and atomic perturbation are provided in the SI. 
	
	\subsection{DFT Parameters}
The DFT\cite{PhysRev.140.A1133, PhysRev.136.B864} calculations were carried out using the Vienna \textit{ab initio} Simulation Package (VASP) \cite{Kresse1996,KRESSE199615}. The exchange-correlation energy was described by the van der Waals density functional \textit{optB88} \cite{optB88}, which was adopted to properly capture dispersion interactions in layered and two-dimensional systems. Collinear spin polarization was considered throughout. A plane-wave basis set with an energy cutoff of 520~eV was used for the valence electrons. Input files were generated automatically using the \texttt{MPStaticSet} class in the \texttt{pymatgen} library. The Brillouin-zone sampling was determined by the \texttt{pymatgen} automatic scheme with a target grid density of 5000 \cite{Ong2013b}, and the sampling along the vacuum direction was limited to one $k$-point. For materials containing strongly correlated \textit{d} or \textit{f} electrons, the DFT+$U$ method \cite{PhysRevB.52.R5467} was employed to treat the on-site Coulomb interaction, with the effective $U$ and exchange $J$ parameters were taken as the default values implemented in \texttt{pymatgen}\cite{Ong2013b}.
	
	\subsection{Training Process}
	
	To ensure reliable evaluation of model performance, the dataset was divided into training, validation, and test sets in a 90\%, 5\%, and 5\% ratio, respectively. For the training of the Uni2D model, the loss function is defined as:
	\begin{equation}
		L = \omega_e \ell(e, e_{\text{DFT}}) + \omega_f \ell(f, f_{\text{DFT}}) + \omega_\sigma \ell(\sigma, \sigma_{\text{DFT}})
	\end{equation}
	The function $\ell(\cdot)$ represents the Huber loss with a parameter $\delta = 0.01$, with $e$ denoting the energy per atom, $f$ the force vector for each atom, and $\sigma$ the stress obtained from the Uni2D model. The hyper-parameters $\omega_e$, $\omega_f$, and $\omega_\sigma$ are assigned values of 1.0, 1.0, and 0.1, respectively. The model is optimized using the Adam optimizer\cite{King014AdamAM}, with an initial learning rate of 0.001 that decays according to a cosine schedule, reaching 1\% of its original value after 100 epochs. The training procedure is performed for 200 epochs with a batch size of 64, utilizing an NVIDIA 4090D GPU for computation.
	
	\section{Results and Discussion}
	
	\subsection{Dataset analysis}
	
	\begin{figure}[htbp]
		\centering
		\includegraphics[width=\textwidth]{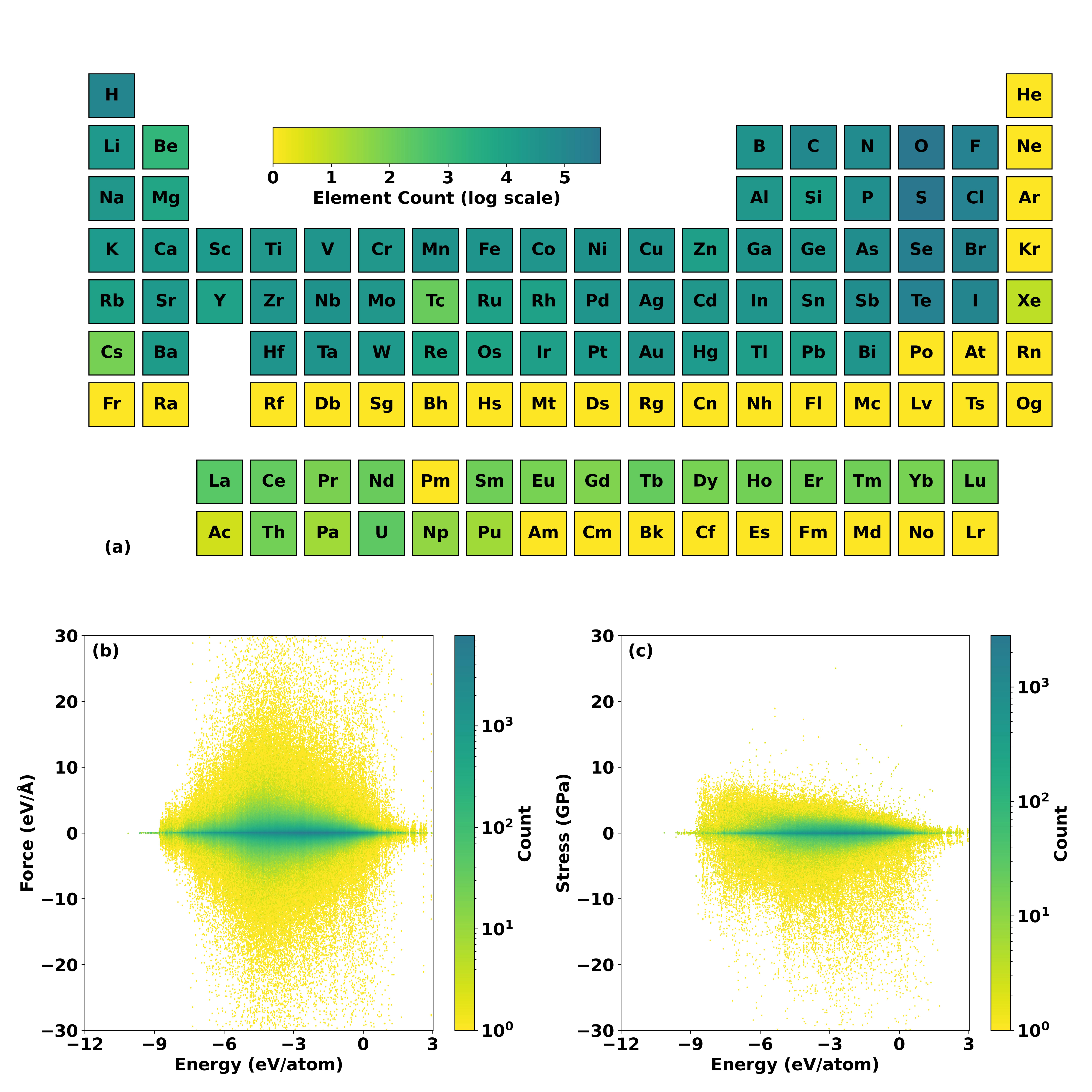}
		\caption{
			Visualization of the dataset used for Uni2D model training. 
			(a) Periodic table colored by the (log-scaled) frequency of elemental occurrences in the dataset. (b) Distribution of atomic force components versus energy per atom.  (c) Distribution of stress components versus energy per atom. 
		}
		\label{fig:table_efs}
	\end{figure}
	
    In this study, we selected approximately 20,000 2D materials and constructed 327,000 mappings between structures and energy, force, and stress. As shown in Figure \ref{fig:table_efs}(a), the elemental distribution reveal that our Uni2D model encompasses 89 elements, with a relatively high proportion of Group N, Group O, and halogen elements, indicating strong predictive capabilities for structures containing these elements. The statistical analysis of the dataset delineates the ranges of energy, force, and stress: the energy per atom ranges from -10.13 to 3.26 eV, the force magnitude ranges from -49.58 to 50.00 eV/\AA, and the stress values ranges from -49.75 to 25.03 GPa (Figure \ref{fig:table_efs}(b) and (c)). Furthermore, as shown in SI Figure~\ref{fig:efs-hist}, the majority of energies (95\%) in the dataset fall within the range from -6.40 to 0.02~eV/atom, while the force components and stress components are predominantly distributed within -2.02 to 2.02~eV/\AA~and -2.18 to 1.57~GPa, respectively.		
    
	\subsection{Model Performance Evaluation}
	A previous study of the M3GNet\cite{m3gnet} bulk-phase model indicated that an IAP model trained solely on energy fails to achieve reasonable accuracy in predicting both forces and stresses (see SI Table \ref{tab:pes_mae_comparison}). Notably, the mean absolute error (MAE) is even higher than the mean absolute deviation of the dataset. This limitation becomes particularly critical in applications involving lattice variations, such as structural relaxation or isothermal-isobaric ensemble molecular dynamics (MD) simulations, where accurate stress predictions are essential. Therefore, in this study, we incorporate energy, force, and stress as training targets and compute the loss function according to the weighting scheme described in the Methods section.
	
\begin{figure}[H]
		\centering
		\includegraphics[width=0.8\textwidth]{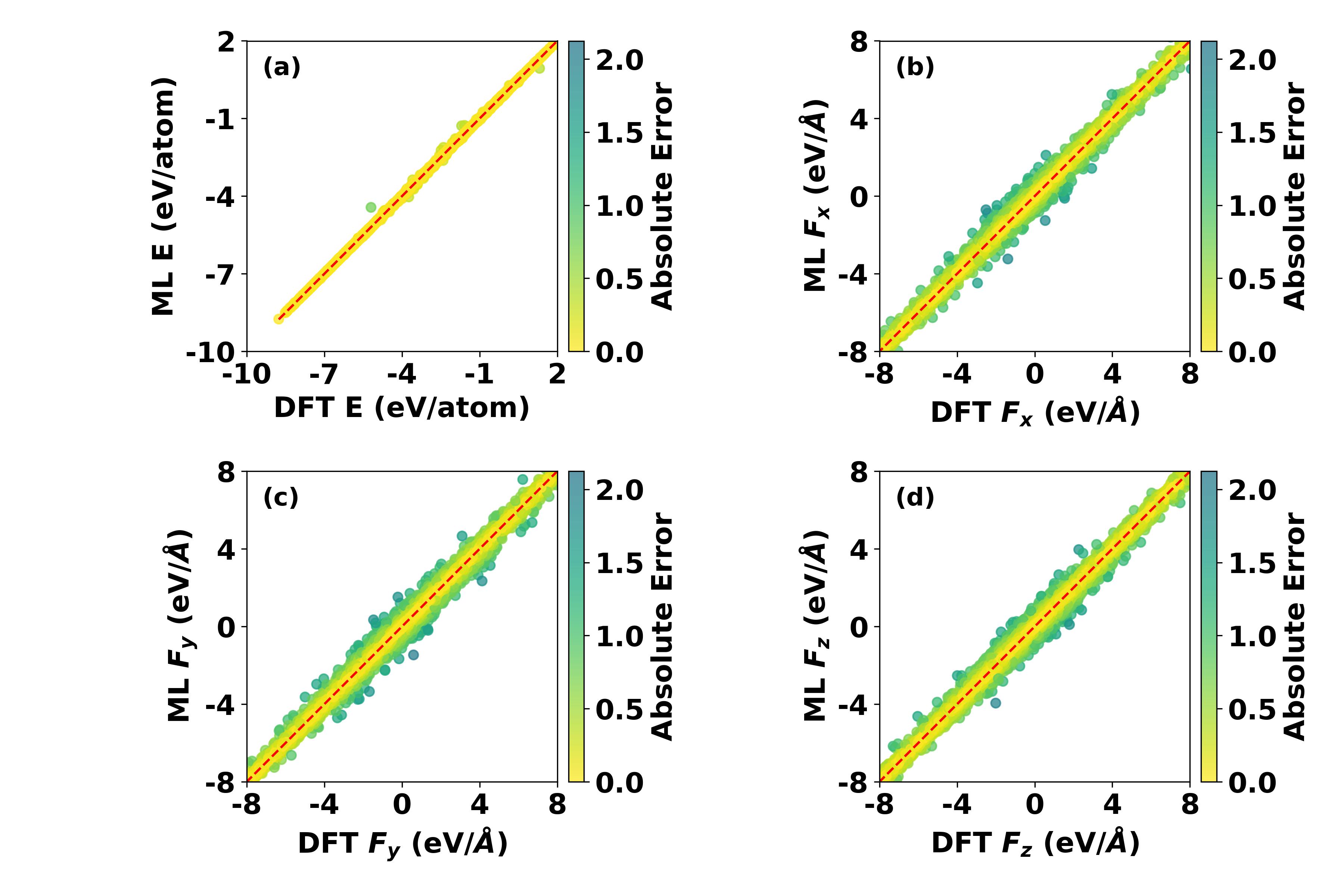}
		\caption{The ML predicted values to the corresponding DFT values for (a) energy (eV/atom) via randomly selected 20000 training structures, (b) force in the x-direction (eV/\AA), (c) force in the y-direction (eV/\AA), and (d) force in the z-direction (eV/\AA). The color bar represents the absolute error between the predicted and true values. The red dashed line indicates perfect agreement \(y = x\).}
		\label{fig:fitting-ef-train}
	\end{figure}

	In Figure~\ref{fig:fitting-ef-train}, we present a comparative analysis between the predictions of the Uni2D and reference values from DFT for energy and force components for the training set. It can be seen that despite the presence of a few outliers with relatively high absolute energy errors, the ML model demonstrates an excellent correlation with the DFT results for the majority of data points, achieving a coefficient of determination of $R^2 = 0.99$ and an MAE of 0.005 eV/atom. In Figure~\ref{fig:fitting-ef-train}(b)–(d), the force predictions are displayed, where $F_x$, $F_y$, and $F_z$ achieve $R^2 = 0.99$, and the corresponding MAEs are 0.052 eV/\AA, 0.056 eV/\AA, and 0.063 eV/\AA, respectively. For the test set, the model maintains a strong linear correlation between predictions and DFT values, further demonstrating its generalization capability (see Figure~\ref{fig:fitting-test-slab}).
	
	Considering the potential similarity between configurations derived from the same parent materials and the risk of overestimating generalization performance, we further evaluated the model on an independently generated dataset. Specifically, we generated 1,685 random 2D structures via PyXtal \cite{pyxtal} and evaluated their energies, forces, and stresses. To assess robustness, four models trained with different random seeds were used for prediction, and the results are summarized in Table~\ref{tab:efs_average}. The averaged MAEs are 0.101 eV/atom for energy, 0.112 eV/\AA\ for force, and 0.152 GPa for stress, respectively. These results provide a more conservative estimate of the model's predictive capability on structurally diverse and previously unseen configurations.
	\begin{figure}[htbp]
	\centering
	\includegraphics[width=0.8\textwidth]{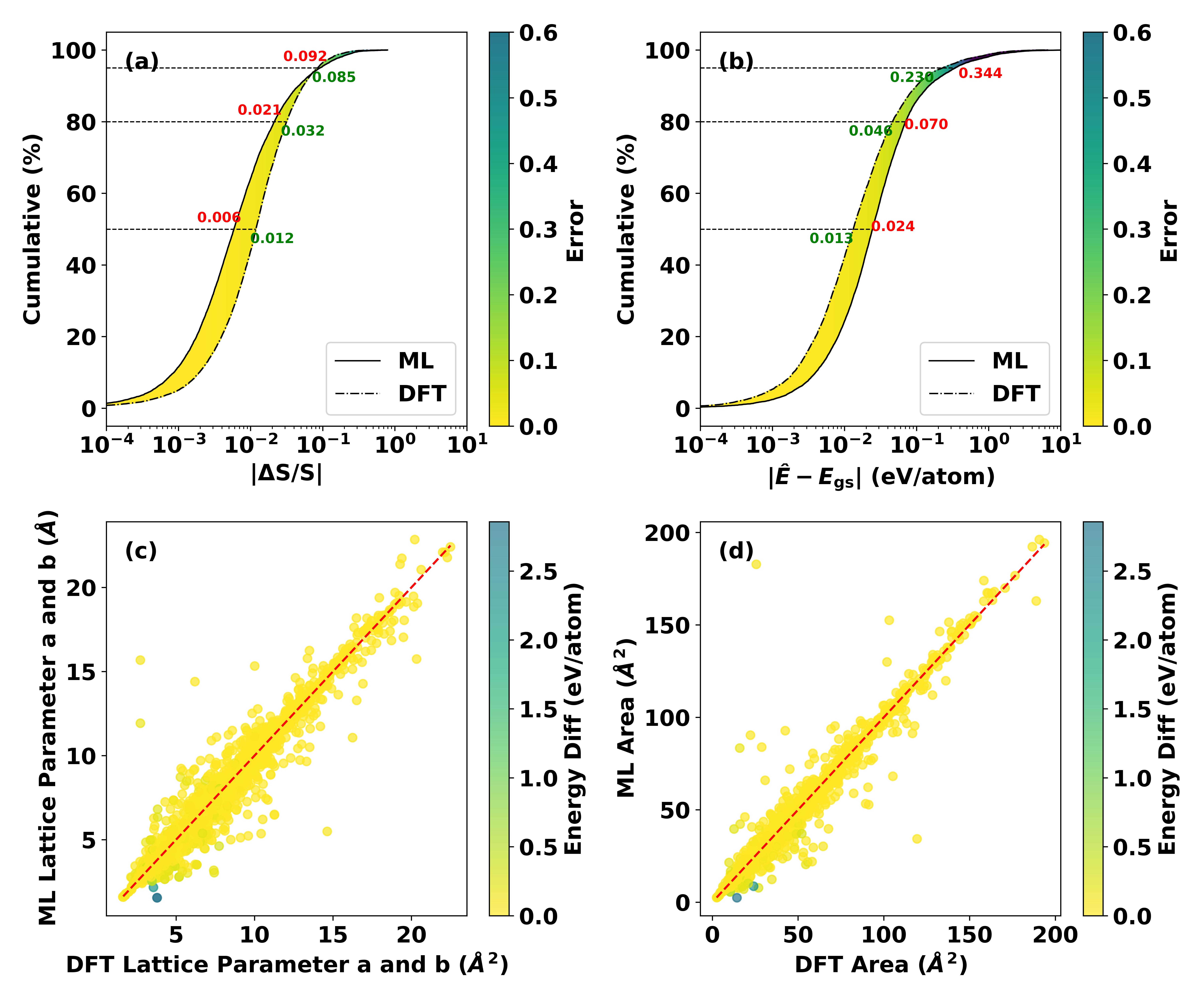}
	\caption{Comparison between DFT and ML predictions for lattice parameters, area errors, and energy errors. (a) The cumulative distribution function (CDF) of the relative area error $\left| \Delta S / S \right|$ between DFT and ML calculations. The shaded area between the curves indicates the error region, colored according to the error value, with annotations marking the 50\%, 80\%, and 95\% cumulative percentages. (b) The CDF of the absolute energy error $\left| \hat{E} - E_{\text{gs}} \right|$ (eV/atom) between DFT and ML calculations. Similarly, the shaded region shows the error difference between DFT and ML, with horizontal lines marking the percentile values. (c) Scatter plot comparing the DFT and ML lattice parameters (\texttt{a} and \texttt{b}) combined, where each data point represents a pair (\texttt{a}, \texttt{b}) from both methods. The data points are colored according to the energy difference (eV/atom), and the diagonal dashed red line represents the perfect agreement line. (d) Scatter plot comparing the DFT and ML calculated areas, similarly colored according to the energy difference. The diagonal line again indicates ideal agreement. Color intensity in the CDF plots (a and b) indicates the magnitude of the relative error, while the scatter plots (c and d) use color to represent energy differences. 
	}
	\label{fig:relaxation}
\end{figure}

   To evaluate the performance of the model in structural optimization, we conducted tests using two datasets. The first dataset consists of structures from the C2DB database, while the second dataset is derived from 2dMatpedia. For the C2DB data, we performed both DFT and ML optimizations. The cumulative distribution of the model errors, as shown in Figure~\ref{fig:relaxation}(a–b), indicates that for 50\% of the ground state structures, the relative area and absolute energy  errors are less than 0.012 and 0.024 eV/atom, respectively. For the second dataset, we directly compared the optimized structures from the database with those obtained from ML optimization. As depicted in Figure~\ref{fig:relaxation}(c–d), the scatter plots demonstrate high linearity between the DFT- and ML-calculated areas and the corresponding lattice parameters, with $R^2$ values of 0.98 and 0.97, respectively. Additionally, the color mapping reveals that the vast majority of structures exhibit low absolute energy errors between the DFT and model predictions, indicating the model's robust performance in predicting lattice properties.

Although no universal IAP model specifically tailored for 2D materials has been proposed to date, the Matbench leaderboard contains numerous PES models primarily developed for three-dimensional bulk systems. To evaluate their generalization capability for 2D systems and enable a fair comparison, we constructed a benchmark dataset comprising approximately 2,000 DFT-calculated structures. For consistency with the training protocols of existing models such as MatterSim, CHGNet, M3GNet, and MACE, the reference energies and structural properties for these models were computed using a PBE (or PBE+U) setup without explicit vdW corrections, while the Uni2D model was evaluated using its native optB88 vdW +U setup. For each model, the predicted initial energy, relaxed energy, and relaxed structural area were compared against their corresponding DFT references, and the performance was quantified using MAE and $R^2$. Under these settings, our model consistently achieved lower MAEs and higher $R^2$ values in both energy and structural predictions, indicating proper accuracy and generalization performance.

\begin{table}[H]
	\centering
	\begin{threeparttable}
		\caption{Comparison of MAE and $R^2$ for initial, relaxed energy and relaxed area based on different ML IAPs.}
		
		\begin{tabular}{lcccccc}
			\toprule
			\multirow{2}{*}{Model} & \multicolumn{2}{c}{Initial Energy (eV/Atom)} & \multicolumn{2}{c}{Relaxed Energy (eV/Atom)} & \multicolumn{2}{c}{Relaxed Area (\AA$^2$)} \\
			\cmidrule{2-3} \cmidrule{4-5} \cmidrule{6-7}
			& MAE & $R^2$ & MAE & $R^2$ & MAE & $R^2$ \\
			\midrule
			Uni2D        & 0.005 & 0.999 & 0.013 & 0.999 & 0.518 & 0.990 \\
			Mattersim    & 0.054 & 0.998 & 0.062 & 0.997 & 0.917 & 0.974 \\
			CHGNet       & 0.295 & 0.958 & 0.331 & 0.945 & 1.806 & 0.933 \\
			M3GNet       & 0.124 & 0.988 & 0.147 & 0.982 & 1.728 & 0.972 \\
			MACE-mh-1    & 0.023 & 0.998 & 0.025 & 0.997 & 0.894 & 0.901 \\
			MACE-mp-0a   & 0.088 & 0.994 & 0.065 & 0.997 & 1.386 & 0.964 \\
			\bottomrule
		\end{tabular}
		
\begin{tablenotes}
	\footnotesize
	\item Note: Each model is evaluated against its own DFT reference (Uni2D: optB88+U; others: PBE+U), without cross-comparison between different DFT levels.
\end{tablenotes}
	\end{threeparttable}
	\label{tab:model-compare}
\end{table}

	In addition, computational efficiency is a critical metric for evaluating model performance. To this end, we systematically compared the computation time required for static calculations using DFT and the ML model on the test dataset. As shown in SI Figure~\ref{fig:dft_vs_ml_time}, DFT static calculations typically require on the order of hundreds of seconds per structure, with an average time of approximately 224 seconds, whereas the ML model completes most static evaluations within 0.2 seconds. Based on the comparison of both average and median computation times, the ML approach achieves an estimated acceleration of approximately $1296.4\times$ (average-based) and $1453\times$ (median-based) relative to DFT. This substantial improvement in computational efficiency highlights the strong potential of the ML model for large-scale, high-throughput static evaluations and accelerated materials discovery.

	\subsection{Property prediction}
		\begin{figure}[htp]
		\centering
		\includegraphics[width=.9\textwidth]{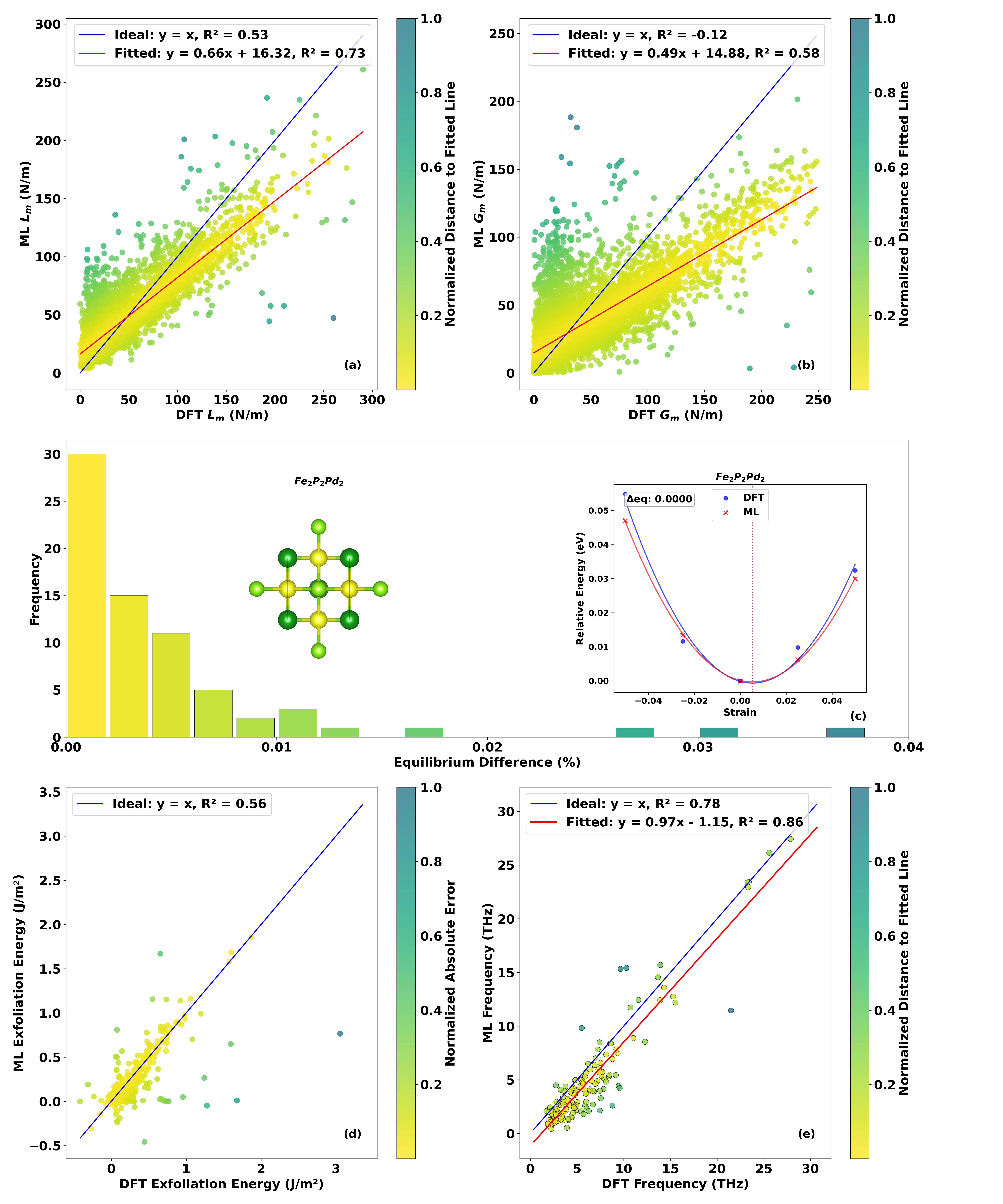}
		\caption{Comparison between ML predictions and DFT calculations for various mechanical and vibrational properties. (a) Layer modulus ($L_m$) predictions versus DFT-calculated values for materials from the C2DB database. (b) Shear modulus ($G_m$) predictions compared to DFT-calculated values, also from the C2DB database. For both (a) and (b), the solid blue line represents the ideal agreement (\(y = x\)), and the red line is a linear fit to the data points. (c) Distribution of equilibrium strain differences between ML-predicted and DFT-calculated equilibrium positions, based on 100 randomly selected structures. (d) Comparison of the exfoliation predictions from ML against DFT-calculated values for materials in the 2dMatpedia dataset, with the ideal (blue) line. (e) Comparison of the average phonon frequency predictions from ML against DFT-calculated values for materials in the MC2D dataset, with the fitted regression line (red) shown alongside the ideal (blue) line.}
		\label{fig:properties}
	\end{figure}
	To evaluate the simulation capability of the proposed universal model for material properties, we performed an in-depth comparative analysis between the model predictions and deformation calculations using data from the C2DB and 2dMatpedia databases. Specifically, we examined the model's accuracy in predicting elastic properties, equations of state, exfoliation energies, and phonon characteristics. 
	
	\paragraph{Mechanical properties}
	As a fundamental property of 2D materials, elasticity~\cite{kiely2021DFTmech, SINGH2021108068} is closely associated with their potential applications. Utilizing a stress-strain-based method, we conducted ML predictions of elastic properties for approximately 7,000 entries from the C2DB database. Specifically, we calculated the effective longitudinal modulus~\cite{molecules28114337}  \(L_m=\frac{1}{4}(C_{11}+C_{22}+C_{12}+C_{21})\) and the shear modulus \(G_m = C_{66}\), as shown in Figure~\ref{fig:properties}(a-b). The linear correlation coefficients between the ML-predicted values of $L_m$ and $G_m$ and the corresponding DFT-calculated values are 0.53 and -0.12, respectively. This indicates a moderate correlation for $L_m$, while virtually no correlation is observed for $G_m$). One possible reason is that elastic constants, being second derivatives of the total energy with respect to strain, are particularly sensitive to error propagation. This is consistent with previous assessments of the M3GNet model~\cite{m3gnet}, although the present model exhibits improved linearity. Moreover, it is important to note that the calculations in the C2DB database do not include vDW corrections. This omission prevents the emergence of an ideal linear relationship of \(y = x\) between ML predictions and DFT references. To better capture the actual relationship, we performed linear regression between the two datasets and found that the fitted lines are not parallel to \(y = x\). After fitting, the correlation coefficients significantly improved to 0.73 for $L_m$ and 0.58 for  $G_m$, respectively. This deviation is attributed to the influence of vDW interactions, particularly in metallic and semiconducting systems, as well as the varying magnitude of such corrections across different material classes. Further analysis based on outlier filtering yields improved correlations, as discussed in the SI (see Figure~\ref{fig:elastic_modulus}).

	\paragraph{Equation of State}
	Unlike equations of state (EOS) for three-dimensional systems\cite{PhysRevB.103.184102}, the EOS of 2D systems is evaluated using a biaxial strain approach. To assess the model's performance, we randomly selected 100 structures from the C2DB database and performed simultaneous EOS calculations using both DFT and ML methods. The statistical analysis, presented as a histogram in Figure~\ref{fig:properties}(c), reveals that the majority of the systems exhibit a relative error in equilibrium position of less than 0.01, demonstrating the model's robust capability in describing the equilibrium state. As an illustrative example, the inset of Figure~\ref{fig:properties}(c) shows the EOS of the \ce{Fe2P2Pd2} structure calculated using both DFT and ML methods, with a deviation of only 0.0013. Additionally, the EOS of nine other randomly selected systems is presented in the SI Figure~\ref{fig:eos_comparison}, further highlighting the consistency between DFT and ML predictions. However, it should be noted that the evaluated structures are drawn from the C2DB database, and some of them may overlap with or resemble configurations in the training dataset, particularly for well-studied layered materials. Therefore, the reported EOS results may, to some extent, reflect interpolation within the learned configuration space rather than fully independent generalization to unseen systems.

	\paragraph{Exfoliation energy}
	Exfoliation energy, a critical property of materials, directly determines whether a material can be exfoliated mechanically, similar to graphene~\cite{yi2015review}. In this study, we employed the optB88 functional, consistent with the exfoliation energy calculations in 2dMatpedia, to perform ML-based simulations for structures with this property, enabling a comparative analysis of exfoliation energies. In 2dMatpedia, the exfoliation energy is defined as \(E_{exf} = E_{2D} - E_{bulk}\), where \(E_{2D}\) and \(E_{bulk}\) represent the total energy per atom of the 2D material and its corresponding bulk parent material, respectively. The scatter plot comparison reveals a good linear relationship between the DFT and ML predictions with a linear correlation coefficient of 0.56 (see Figure~\ref{fig:properties}(d)). However, for certain specific systems, significant discrepancies in exfoliation energy were observed, which may result from inaccuracies in bulk structure energy predictions. This issue likely arises because the current model incorporates only a limited number of bulk structures, leading to potential errors when calculating \(E_{bulk}\).

	\paragraph{Phonon simulation}
	To evaluate the capability of the Uni2D model in describing lattice dynamics, we performed phonon simulations for 180 structures from the MC2D database~\cite{Campi2023_MaterialsCloud2D}. The average phonon frequency was calculated using 
	\begin{equation}
		\bar{\omega} = \frac{\int \omega \, g(\omega) \, d\omega}{\int g(\omega) \, d\omega}.
	\end{equation}
	The results show a clear linear correlation between ML and DFT values, with an $R^2$ value of 0.78. Similar to the elasticity results discussed above, a systematic deviation from the ideal \(y=x\) relationship is observed, which may arise from differences in exchange–correlation treatment between the reference DFT data and the ML model, as illustrated by the fitted line in Figure~\ref{fig:properties}(e). As an additional validation, we further evaluated a series of 2D auxetic \ce{M4X8} materials~\cite{wang2025M4X8}, where the predicted average phonon frequencies also exhibit a strong linear relationship between ML and DFT results, with an $R^2$ value of 0.84 (see SI Figure~\ref{fig:M4X8_dft_ml_comparison}). Moreover, we performed additional DFT calculations at the same level of accuracy for selected systems, including \ce{TiGe2As4}, \ce{MoC2}, \ce{Mo2Se4}, and graphene, and found that the phonon properties predicted by the Uni2D model are in good agreement with the corresponding DFT results, further supporting the model's capability for lattice dynamics simulations (see SI Figure~\ref{fig:phonon_comparison}).
	
	\subsection{Application Cases in 2D Materials}
	
	\paragraph{Activation Energy of \ce{Li} Diffusion in \ce{MoS2} Bilayers}

	Lithium-ion batteries (LIBs) are highly efficient energy storage devices characterized by high energy density, compact size, long cycle life, and low cost\cite{doi:10.1021/ja3091438,C2JM14305D}. An ideal electrode material should not only exhibit suitable intercalation/deintercalation voltages but also possess good lithium-ion mobility\cite{B901825E,doi:10.1021/cr020731c}. Owing to its unique layered structure, \ce{MoS2} enables reversible intercalation and deintercalation of \ce{Li+} and is therefore widely regarded as a promising electrode material for lithium-ion batteries\cite{https://doi.org/10.1002/adfm.201404078,CUI201822,doi:10.1021/acsami.7b00248}.
	    \begin{figure}[H]
		\centering
		\includegraphics[width=0.6\textwidth]{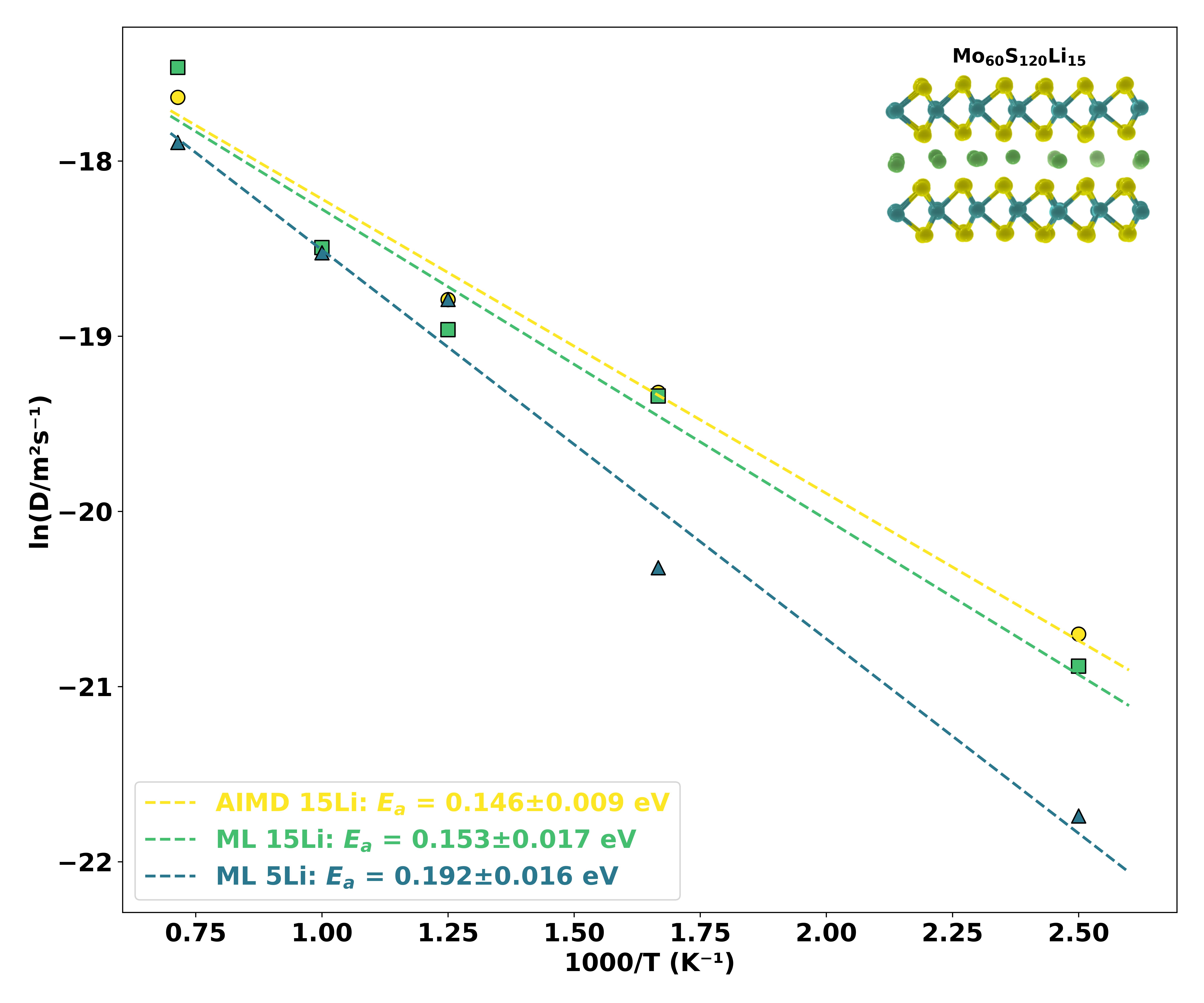}
		\caption{
			Arrhenius plots of the natural logarithm of the diffusion coefficient \(\ln(D)\) versus inverse temperature \(1000/T\) for \ce{Li+} diffusion in \ce{Mo60S120Li15} and \ce{Mo60S120Li20} systems. The dashed lines represent linear fits to AIMD and ML data, with corresponding activation energies (\(E_a\)) extracted from the slopes. The inset image illustrates the atomic structure of the \ce{Mo60S120Li15} system used in the simulations.
		}
		\label{fig:li-diffusion}
	\end{figure}
	
	To evaluate \ce{Li+} diffusion in \ce{MoS2} systems, the self-diffusion coefficient was calculated from atomic trajectories obtained from MD simulations, and the diffusion activation energy was extracted using the Arrhenius relation. As shown in Figure~\ref{fig:li-diffusion}, we investigated two representative systems, \ce{Mo60S120Li15} and \ce{Mo60S120Li20}, and compared the results with traditional AIMD simulations. The Uni2D model yields activation energies in good agreement with AIMD results. Specifically, the diffusion barrier for \ce{Mo60S120Li15} is 0.153 $\pm$ 0.017 eV, which is close to the AIMD value of 0.146 $\pm$ 0.009 eV, while the \ce{Mo60S120Li20} system exhibits a higher diffusion barrier of 0.192 $\pm$ 0.016 eV. These results indicate that increasing lithium concentration leads to higher diffusion barriers and reduced lithium-ion mobility. Further details can be found in the SI.

	\paragraph{2D Material Searching}
	
    \begin{figure}[htbp]
		\centering
		\includegraphics[width=1.0\textwidth]{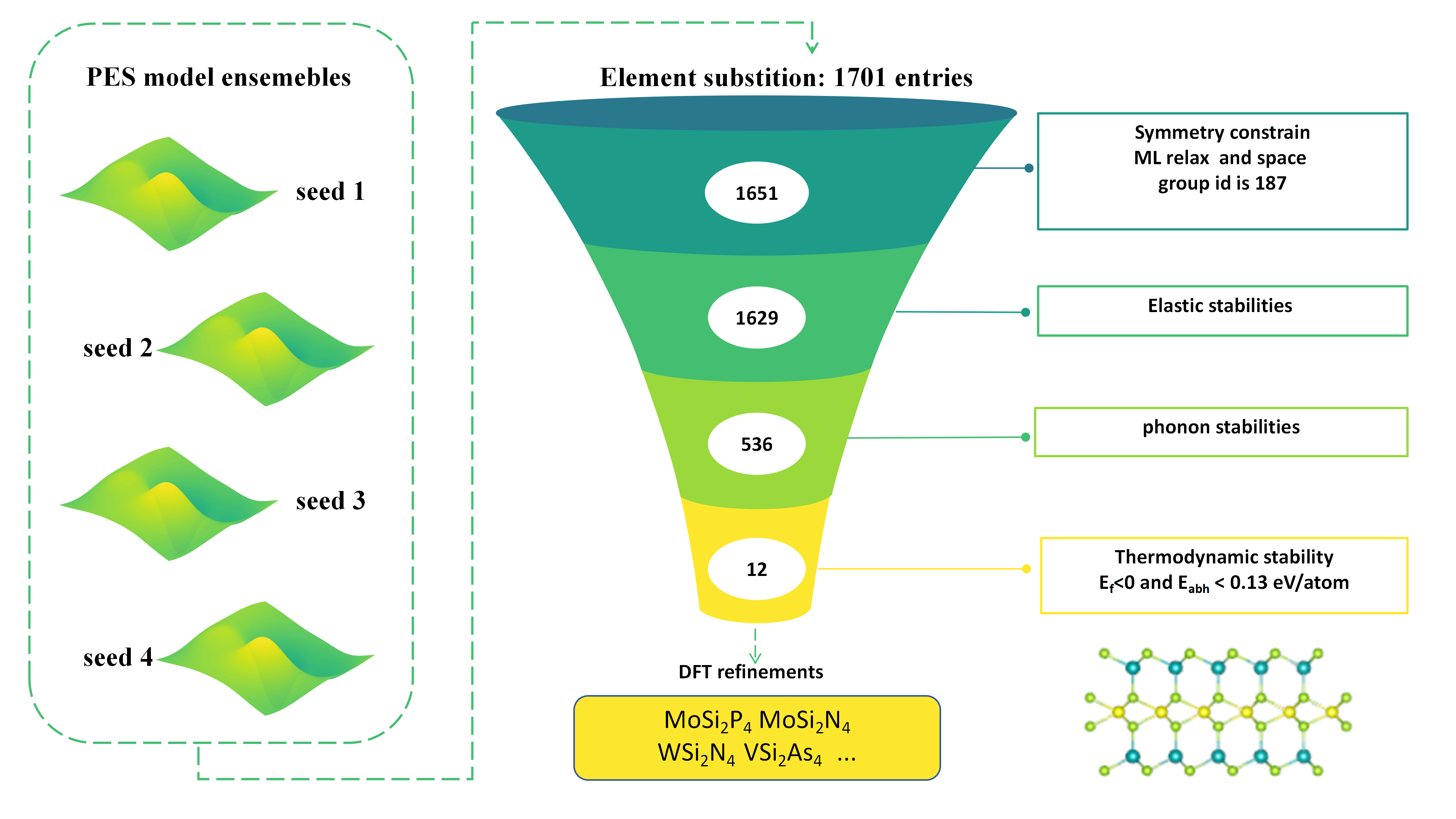}
		\caption{Schematic illustration of zero-shot application of the universal PES model for the \ce{MA2Z4} system and high-throughput materials discovery.}
		\label{fig:agent}
	\end{figure}

To further evaluate the transferability and generalization capability of the Uni2D model, we adopted a zero-shot strategy, in which the base model was directly applied to materials screening and discovery without any additional fine-tuning. As a representative test case, we selected the \ce{MA2Z4} structure, a recently proposed family of vDW layered materials\cite{Zhou2025}. Owing to its distinctive layered structure and compositional flexibility, this material family exhibits diverse properties and promising applications in electronics, catalysis, and energy-related fields  \cite{adfm.202214050,s41467-021-22324-8,MoS2N4-science,Zhen2024eem}. Based on this structural prototype, we systematically generated 1,701 isostructural candidates by substituting A- and Z-site atoms with elements from the same groups and replacing the transition-metal element at the M site.
  \begin{table*}[htbp]
	\centering
	\caption{Lattice parameter $a$ (\AA), formation energy $E_f$ (eV/atom), energy above the convex hull $E_{abh}$ (eV/atom), HSE06 band gap (eV), and references of the screened compounds. The lattice parameters, $E_f$, and $E_{abh}$ are computed using the MPStatic workflow with default GGA/GGA+$U$ settings to ensure consistency with the Materials Project database. The band gaps are calculated using the HSE06 functional, and the total band gap is reported for spin-polarized systems.}
	\label{tab:screened_compounds}
	\begin{tabular}{lccccc}
		\hline
		formula & $a$ & $E_f$ & $E_{abh}$ & gap & refs\\
		\hline
		MoSi$_{2}$P$_{4}$   & 3.470 & -0.303 & 0.000 & 0.98 &  \cite{Kumavat_2024}\\
		WSi$_{2}$P$_{4}$    & 3.477 & -0.240 & 0.000 & 0.80 &  \cite{s41467-021-22324-8}\\
		MoSi$_{2}$N$_{4}$   & 2.908 & -0.957 & 0.069 & 2.29 & \cite{MoS2N4-science} \\
		VSi$_{2}$P$_{4}$    & 3.488 & -0.244 & 0.095 & 0.71 & \cite{s41467-021-22324-8}\\
		TiSi$_{2}$As$_{4}$  & 3.714 & -0.215 & 0.099 & 1.00 & \cite{Verzola2024MA2Z4}\\
		TiSi$_{2}$P$_{4}$   & 3.557 & -0.327 & 0.103 & 1.04 & \cite{Verzola2024MA2Z4}\\
		VSi$_{2}$As$_{4}$   & 3.645 & -0.102 & 0.115 & 0.34 & \cite{PhysRevB.106.235401} \\
		MoGe$_{2}$N$_{4}$   & 3.020 & -0.209 & 0.115 & 1.36 & \cite{zhang2022thermoelectric}\\
		HfSi$_{2}$P$_{4}$   & 3.635 & -0.356 & 0.118 & 0.78 & \cite{Verzola2024MA2Z4} \\
		HfSi$_{2}$As$_{4}$  & 3.793 & -0.241 & 0.123 & 0.72 & \cite{Verzola2024MA2Z4}\\
		ZrSi$_{2}$P$_{4}$   & 3.640 & -0.365 & 0.124 & 0.81 & \cite{Verzola2024MA2Z4}\\
		WSi$_{2}$N$_{4}$    & 2.915 & -0.944 & 0.128 & 2.66 & \cite{MoS2N4-science}\\
		\hline
	\end{tabular}
	
\end{table*}

Using an ensemble strategy similar to \texttt{DP-GEN}\cite{ZHANG2020107206}, we performed high-throughput initial relaxation of all 1,701 candidate structures with Uni2D. After filtering based on space-group symmetry, 1,651 structures were retained on average across the committee models (see SI Table~\ref{tab:model_compare}). These structures were subsequently evaluated for mechanical and dynamical stability by computing elastic constants and phonon dispersion relations. Only those satisfying the Born stability criteria \cite{Mazdziarz2019} and exhibiting no imaginary phonon modes were retained, resulting in 536 stable candidates. To further assess thermodynamic stability, these 536 structures were fully relaxed using DFT calculations. Thermodynamic screening was then performed by requiring negative formation energies and energies above the convex hull less than 0.13 eV/atom~\cite{research2022}, yielding 12 potentially synthesizable structures. Phonon calculations at the DFT level were subsequently conducted to confirm their dynamical stability, and ultimately 12 fully stable compounds were identified (see Table~\ref{tab:screened_compounds}, SI Figure~\ref{fig:ma2z4_band_structures} and Figure~\ref{fig:ma2z4_phonon_bands}). Notably, \ce{MoSi2N4} and \ce{WSi2N4} have already been experimentally synthesized \cite{MoS2N4-science}, while \ce{VSi2As4} has been theoretically predicted as a bipolar magnetic semiconductor \cite{wang_prm,research2022} with a high Curie temperature (900 K) \cite{PhysRevB.106.235401}.

Although the present search was constrained to a fixed structural template, the methodology can, in principle, be extended to other structural prototypes and chemical spaces. Notably, the screening results obtained in this work are generally consistent with previous experimental and theoretical studies, supporting the validity of the approach. These findings suggest that the proposed strategy may be useful for large-scale and composition-wide screening, and could facilitate data-driven exploration of functional materials.

	\subsection{LLM-Based Interface for Uni2D}
	
	\begin{figure}[ht]
		\centering
		\includegraphics[width=1.0\textwidth]{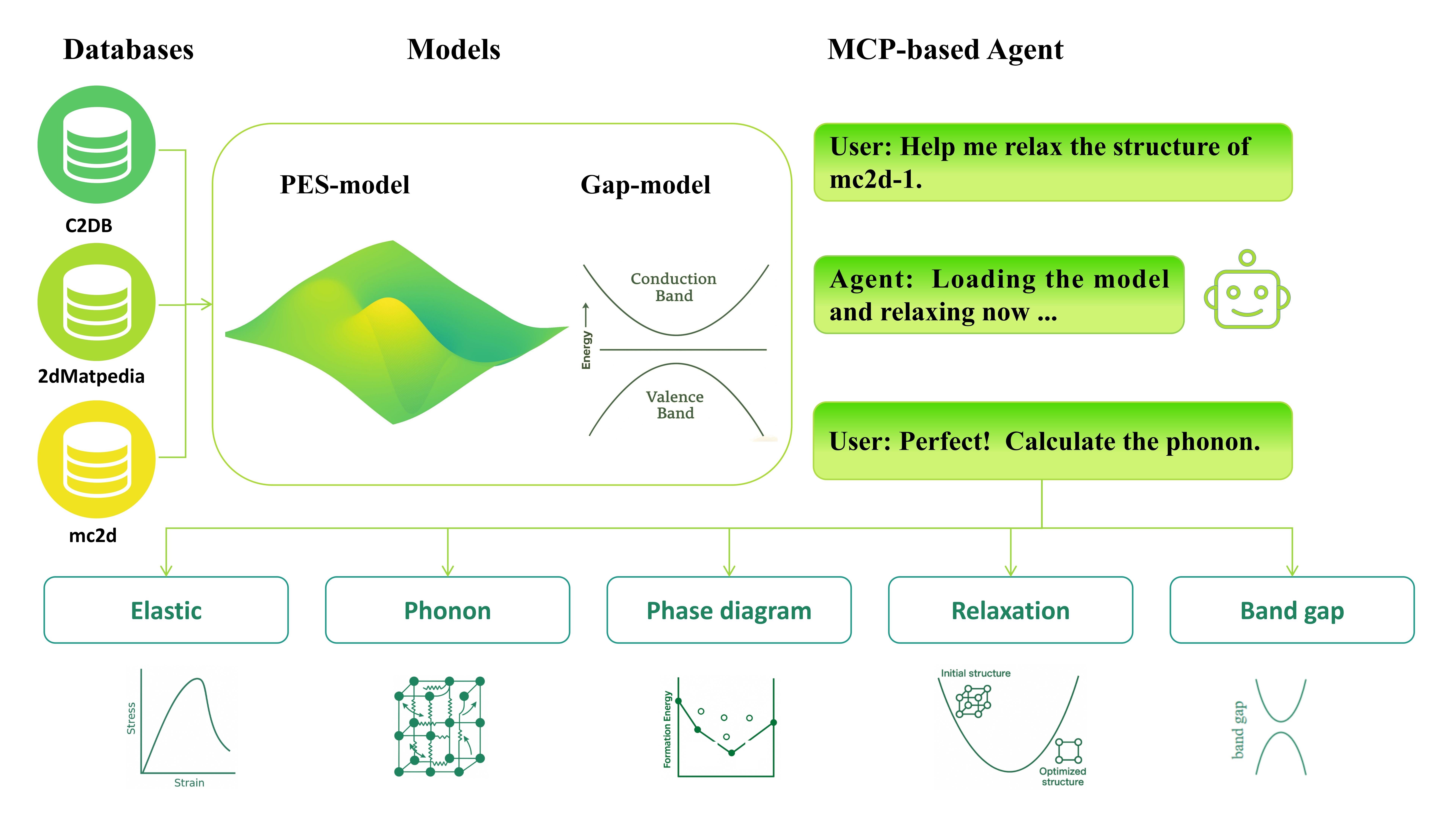}
		\caption{ Schematic illustration of Uni2D  model integrated with an LLM-powered agent for 2D material analysis.}
		\label{fig:llm-agent}
	\end{figure}

To further improve the usability and operational efficiency of the proposed Uni2D model for 2D materials, we integrated it into an autonomous agent system driven by the DeepSeek LLM\cite{deepseek} and built on the UniMatSim framework~\cite{unimatsim}, which enables secure and standardized communication between AI applications and external computational tools and data sources. Within this framework, materials science computational tools, database resources, and analysis functionalities are exposed through unified interfaces, allowing the LLM to seamlessly access and coordinate specialized scientific workflows. Complex tasks, including structure search, crystal structure analysis, and physical property calculations, are encapsulated as modular functions, enabling the agent to autonomously perform simulations such as structure optimization, elasticity evaluation, phonon dispersion analysis, and molecular dynamics simulations (see Figure \ref{fig:llm-agent}). To further evaluate electronic properties, we incorporated GNN-based machine learning models\cite{PhysRevLett.120.145301}, including a classifier for distinguishing metals from semiconductors and a regression model for predicting semiconductor band gaps. By combining Uni2D with these electronic property prediction models and UniMatSim architecture, the system supports natural language-driven human–machine interaction for automated retrieval, analysis, and simulation of 2D materials without requiring extensive computational expertise. Overall, this integrated framework enhances the accessibility, automation, and efficiency of computational materials discovery while maintaining extensibility and interoperability across diverse research workflows. Further details can be found in the SI.

	\section{Conclusion}
		
	In summary, we developed a universal PES model for 2D materials based on the GNN framework MatterSim. By incorporating an enhanced sampling strategy, the proposed model demonstrates strong generalization capabilities across a diverse range of 2D material systems and accurately captures atomic-scale interactions. Owing to its high computational efficiency and predictive accuracy, this model provides a useful tool for high-fidelity simulations of 2D materials, supporting applications such as structure optimization, molecular dynamics, lattice dynamics, and elastic property calculations. Furthermore, when integrated with our LLM-driven intelligent interface, the model exhibits improved usability and accessibility, paving the way for broad applications in inverse materials design of 2D systems.
	
	Despite the promising performance in terms of generality and efficiency, several challenges remain to be addressed: (1) The pursuit of universality introduces a trade-off with accuracy. Enhancing predictive performance will require sampling from a more expansive and chemically diverse configuration space; (2) The current model lacks the capability to predict magnetic properties—such as magnetic moments—which are crucial for studying magnetic 2D materials. In future work, we aim to integrate features of CHGNet or optimize our model based on it, to enable accurate modeling of 2D magnetic systems; (3) The training data primarily consists of DFT-calculated structures under low-temperature conditions, which may limit the model’s reliability in high-temperature or far-from-equilibrium regimes. Incorporating high-temperature configurations will be necessary to extend the applicability of the potential energy surface; (4) as an IAP, the model inherently cannot describe electronic structure, a key aspect of 2D materials functionality. To address this, we envision combining the potential energy model with Hamiltonian-based approaches\cite{Li2022, Tang2024} to establish a truly end-to-end machine learning framework capable of simultaneously predicting both atomic interactions and electronic properties with high efficiency and fidelity.
		
	\begin{acknowledgement}
		
This work is supported by the National Natural Science Foundation of China (Grant No. 22203026), the National Key R\&D Program of China (Grant No. 2022YFA1602601), the National Natural Science Foundation of China (Grant No. 22288201, 12174080, 22403024 and 22203025), the Innovation Program for Quantum Science and Technology (Grant No. 2021ZD0303306), the Strategic Priority Research Program of the Chinese Academy of Sciences (Grant No. XDB0450101), the Fundamental Research Funds for the Central Universities (Grant No. JZ2024HGTB0162) and the Anhui Provincial Natural Science Foundation (2308085QB52) and the Open Research Fund of State Key Laboratory of Precision and Intelligent Chemistry. The computation is performed on the HPC platform of Hefei University of Technology and the University of Science and Technology of China. 
		
	\end{acknowledgement}
		
\section*{Data Availability}
The Uni2D model and Gap model weights, inference scripts, and the LLM-based interface code are publicly available at \url{https://gitee.com/haidi-hfut/Uni2D}, \url{https://gitee.com/haidi-hfut/gap2d}, and \url{https://gitee.com/haidi-hfut/matagent}. 

	\section*{Declarations}
	During manuscript preparation, we utilized GPT-5 LLM to enhance sentence structure, readability, and coherence. We also used Claude 4.6 LLM for optimizing the Python plotting and simulation scripts.
    
	\footnotesize{
	\bibliography{main}
	}

\end{document}



		\vspace{3ex}
		\renewcommand{\thefigure}{S\arabic{figure}}
		\setcounter{figure}{0}
		\renewcommand{\thetable}{S\arabic{table}}
		\setcounter{table}{0}

		\section{Model Architecture}	
		In the M3GNet model, each material structure is represented as a graph $\mathcal{G} = (\mathcal{V}, \mathcal{E}, \mathcal{X}, [\mathbf{M}, \mathbf{u}])$, where:
		
		\begin{itemize}
			\item \textbf{Nodes} $\mathcal{V}$ represent atoms, and each node $i$ contains an atomic feature vector $\mathbf{v}_i$.
			\item \textbf{Edges} $\mathcal{E}$ represent bonds between atoms, with each edge $e_{ij}$ corresponding to the bond feature between atoms $i$ and $j$.
			\item \textbf{Coordinates} $\mathcal{X}$ store the atomic positions $\mathbf{x}_i$ for each atom $i$.
		\end{itemize}
		
		For crystalline materials, the 3$\times$3 lattice matrix $\mathbf{M}$ may be included, along with the global state $\mathbf{u}$ to account for properties such as temperature and pressure.
		
		A key feature of M3GNet is the incorporation of three-body angular interactions, which are modeled using spherical Bessel functions and spherical harmonics. Let $\theta_{jik}$ denote the angle between bonds $e_{ij}$ and $e_{ik}$, then the updated equation $\mathbf{e}'_{ij}$  can be rewritten as:
		
		\begin{equation}
			\mathbf{e}_{ij} = \sum_l j_l\left(z_l \frac{r_{ik}}{r_c}\right) Y_l^0(\theta_{ijk}) \otimes \sigma(W_k \mathbf{v}_k + \mathbf{b}_k) f_c(r_{ij}) f_c(r_{ik})
		\end{equation}
		\begin{equation}
			\mathbf{e}'_{ij} = \mathbf{e}_{ij} + g\left(\tilde{W}_2 \tilde{\mathbf{e}}_{ij} + \tilde{\mathbf{b}}_2 \right) \otimes \sigma\left(\tilde{W}_1 \tilde{\mathbf{e}}_{ij} + \tilde{\mathbf{b}}_1 \right)
		\end{equation}
		where $j_l$ is the spherical Bessel function evaluated at roots $z_l$,  $r_c$ is the cutoff radius, $Y_l^0(\theta_{ijk})$ is the spherical harmonics function with $m=0$,  $\otimes$ denotes the element-wise product, $\sigma(\cdot)$ is the sigmoid activation function, and $f_c(r)$ is the smooth cutoff function defined as: 
		
		\begin{equation}
			f_c(r) = 1 - 6\left(\frac{r}{r_c}\right)^5 + 15\left(\frac{r}{r_c}\right)^4 - 10\left(\frac{r}{r_c}\right)^3
		\end{equation}
		The $g(x) = \sigma(x)$ is a nonlinear activation function, and $\tilde{W}$ and $\tilde{\mathbf{b}}$ represent learnable weight matrices and bias vectors. The indices $l = 0, 1, \dots, l_{\max} - 1$ and $n = 0, 1, \dots, n_{\max} - 1$ define the angular and radial components of the feature vector $\mathbf{e}_{ij}$, where $l_{\max}$ and $n_{\max}$ are hyper-parameters that govern the complexity of the angular and radial components.
		
		The resulting atom and bond features $\mathbf{v}_i$ and $\mathbf{e}_{ij}$ are passed to a gated multi-layer perceptron to obtain the prediction of energies, namely: 
		
		\begin{itemize}
			\item \textbf{Energy} $E$: The total energy is $E=\sum{E_i}$, where is $E_i$ is the energy of $i$-th atom in the system.
			\item \textbf{Forces} $\mathbf{f}$: Computed via automatic differentiation as $\mathbf{f} = -\dfrac{\partial E}{\partial \mathbf{x}}$, where $\mathbf{x}$ represents atomic positions.
			\item \textbf{Stress} $\sigma$: Computed as $\sigma = \dfrac{1}{V} \dfrac{\partial E}{\partial \varepsilon}$, where $V$ is the system volume and $\varepsilon$ is the strain.
		\end{itemize}
		
		\section{Data Preparation}
		The augmentation strategy is outlined as follows:
		
		\paragraph{Lattice Strain} Each 2D structure’s lattice is subjected to uniform scaling and shear deformation to introduce different types of strain. The adjustment of the lattice vectors \(\mathbf{L}\) is achieved by applying a deformation matrix \(\mathbf{D}\), which simulates various strain conditions. The lattice transformation is given by the following equation:
		\begin{equation}
			\mathbf{L}_{\text{new}} = \mathbf{L}_{\text{orig}} \times (\mathbf{I} + \mathbf{D})
		\end{equation}
		where \(\mathbf{I}\) is the identity matrix, and \(\mathbf{D}\) is the \(3 \times 3\) deformation matrix that represents the tensile/compressive and shear strains. The specific form of the deformation matrix \(\mathbf{D}\) is as follows:
		\begin{equation}
			\mathbf{D} = 
			\begin{pmatrix}
				r_{xx} & s_{xy} & 0 \\
				s_{yx} & r_{yy} & 0 \\
				0 & 0 & 1
			\end{pmatrix}
		\end{equation}
		Here, \(r_{xx}\) and \(r_{yy}\) represent the tensile or compressive factors along the \(x\)- and \(y\)-directions, respectively, and \(s_{xy}\) and \(s_{yx}\) account for the shear strain. The tensile/compressive factors \(r_{xx}, r_{yy}\) are chosen from the range \([-0.04, 0.04]\), while the shear factors \(s_{xy}, s_{yx}\) are selected from the range \([0.01, 0.02]\), controlling the lattice shape deformation.
		
		\paragraph{Atomic Perturbations} To simulate defects, thermal vibrations, and other physical factors that deviate from the equilibrium structure, we introduce random perturbations to the atomic positions. The perturbation matrix is drawn from a normal distribution, with the mean (\(\mu\)) and standard deviation (\(\sigma\)) controlling the magnitude of the perturbations. The equation is given by:
		\begin{equation}
			\mathbf{R}_{\text{new}} = \mathbf{R}_{\text{orig}} + \mathbf{P}
		\end{equation}
		where \(\mathbf{P} \sim \mathcal{N}(\mu, \sigma)\) represents the perturbations applied to the Cartesian coordinates. In this study, we choose \(\mu = 0.1 \, \text{\AA}\) and \(\sigma = 0.05 \, \text{\AA}\) to ensure that the changes in atomic positions are physically reasonable.
		
		In addition, the training dataset includes approximately 16,000 energy-force-stress mapping entries corresponding to bulk-phase structures. These data originate from two main sources. One part is derived from the OQMD database\cite{Kirklin2015OQMD,Saal2013}, which contains unary to ternary bulk compounds primarily used for constructing phase diagrams of two-dimensional materials in the C2DB database. The other part consists of ground-state stable structures retrieved via the Materials Project API\cite{10.1063/1.4812323, ONG2015209}. By incorporating this subset of stable bulk-phase structures, our aim is twofold: to improve the robustness of the model's predictions and to enhance its ability to learn the features of stable configurations, thereby providing strong data support for the accurate prediction of 2D material stability.
		
		\newpage
		\section{Statistic Analysis}
		
\begin{figure}[H]
			\centering
			\includegraphics[width=\textwidth]{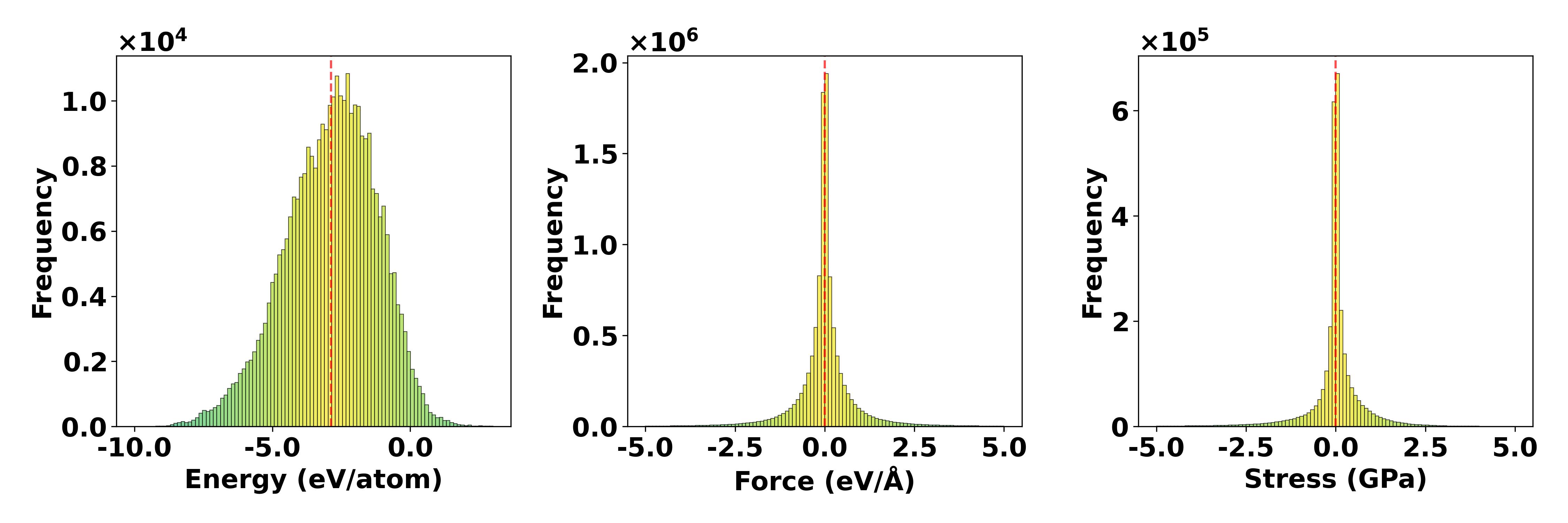}
			\caption{
				Distributions of energy, force, and stress in the training dataset. 
				Energy is given in eV/atom, force in eV/\AA, and stress in GPa. 
			}
			\label{fig:efs-hist}
		\end{figure}

		Figure~\ref{fig:efs-hist} shows the distributions of energy, force, and stress in the training dataset, illustrating the overall data coverage and value ranges used during model training.
		
		Table~\ref{tab:pes_mae_comparison} compares the prediction accuracy of PES models trained with different combinations of energy (E), forces (F), and stress (S). Training with energy only results in large force and stress errors, while including force labels significantly improves force prediction and reduces stress error. Further incorporating stress labels leads to additional improvement in stress accuracy while maintaining similar energy and force performance. These results indicate that jointly training with energy, forces, and stress provides the most balanced and accurate description of the potential energy surface, and therefore this training strategy is adopted for the final Uni2D model.
		\begin{table}[htbp]
			\centering
			\begin{tabular}{cccc}
				\hline
				\textbf{Metric (MAE)} & \textbf{E} & \textbf{E, F} & \textbf{E, F, S} \\
				\hline
				Energy (eV/atom) & 0.007 & 0.007 & 0.006 \\
				$F_x$ (eV/\AA)& 0.190 & 0.055 & 0.055 \\
				$F_y$ (eV/\AA)& 0.217 & 0.061 & 0.061 \\
				$F_z$ (eV/\AA)& 0.272 & 0.068 & 0.068 \\
				Stress (GPa) & 0.218 & 0.122 & 0.082 \\
				\hline
			\end{tabular}
			\caption{Mean absolute error comparison for PES models trained with different combinations of energy (E), forces (F), and stress (S).}
			\label{tab:pes_mae_comparison}
		\end{table}

\begin{figure}[htbp]
			\centering
			\includegraphics[width=.8\textwidth]{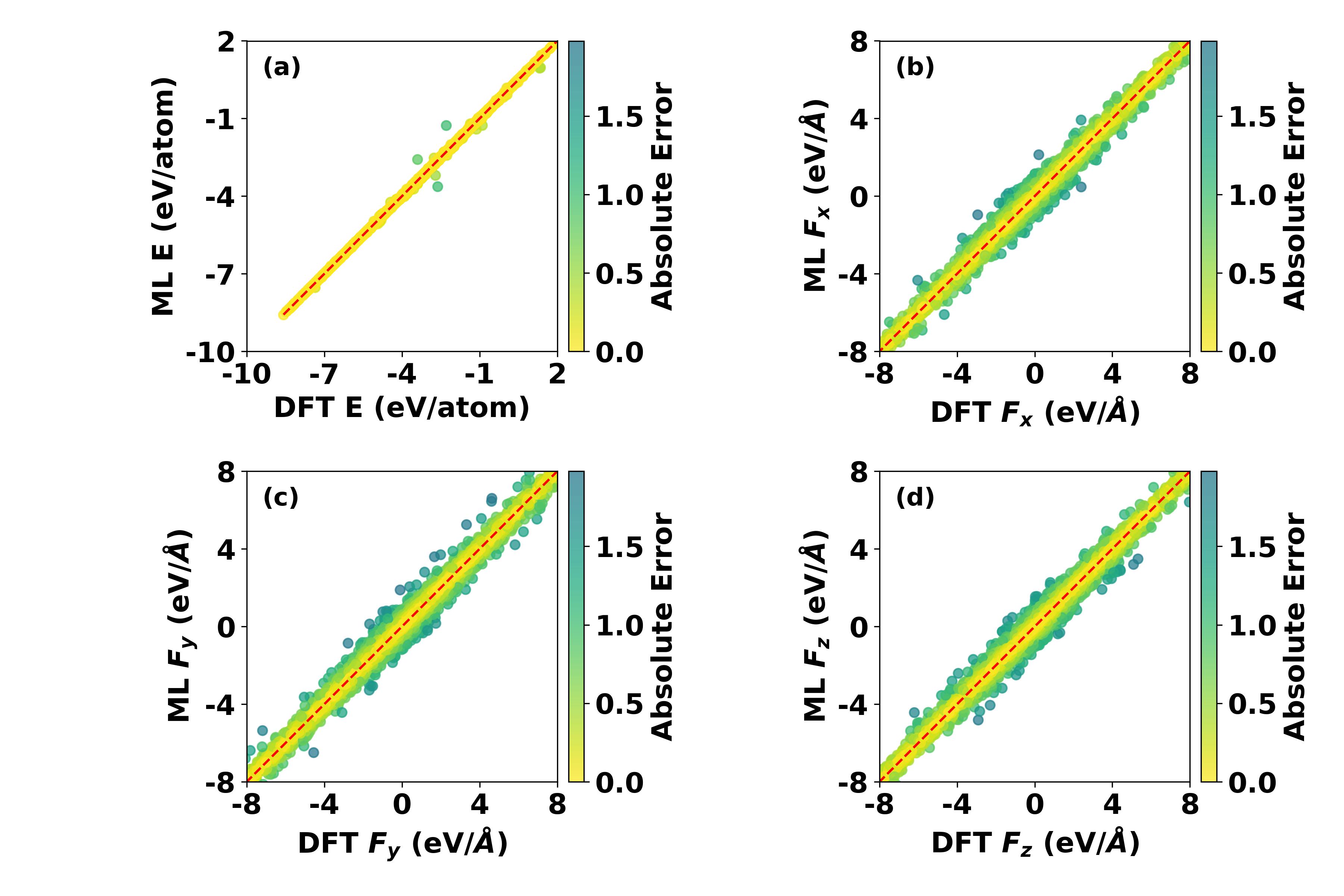}
			\caption{The ML predicted values to the corresponding DFT values for (a) energy (eV/atom) structures of test dataset, (b) force in the x-direction (eV/\AA), (c) force in the y-direction (eV/\AA), and (d) force in the z-direction (eV/\AA). The color bar represents the absolute error between the predicted and true values. The red dashed line indicates perfect agreement (y = x)}
			\label{fig:fitting-test-slab}
		\end{figure}

Figure~\ref{fig:fitting-test-slab} presents the correlation between ML predictions and DFT reference values for energy and force components in the test dataset. It should be noted that this test set is obtained from a random split of the augmented dataset. Due to the structural similarity among configurations derived from the same parent materials, this evaluation may lead to an optimistic estimate of the model’s predictive performance.

To further assess the robustness and generalization capability of the Uni2D model, we evaluated its performance on an independently generated dataset consisting of 1,685 random 2D structures created using PyXtal \cite{pyxtal}. These structures are structurally diverse and not derived from the training dataset, thereby providing a more stringent test of model transferability. Energies, forces, and stresses were predicted using four models trained with different random seeds to evaluate prediction consistency and robustness.

Table~\ref{tab:efs_average} summarizes the prediction performance of models trained with different random seeds. Overall, the results show consistent predictive accuracy across different models, with only minor variations in performance. The averaged mean absolute errors (MAEs) are 0.101~eV/atom for energy, 0.112~eV/\AA\ for force, and 0.152~GPa for stress. These results indicate that the Uni2D model exhibits stable performance across different training initializations and maintains reliable predictive capability on structurally diverse and previously unseen configurations. This independent test provides a more rigorous and reliable evaluation of the model’s generalization capability across structurally diverse systems.

\begin{table}[htbp]
	\centering
	\caption{Performance comparison of four models trained with different random seeds on energy, force, and stress prediction evaluated on the test set.}
	\begin{tabular}{lccc}
		\hline
		\textbf{Model} & \textbf{$E_{\mathrm{MAE}}$ (eV/atom)} & \textbf{$F_{\mathrm{MAE}}$ (eV/\AA)} & \textbf{$S_{\mathrm{MAE}}$ (GPa)} \\
		\hline
		model-seed1 & 0.122 & 0.113 & 0.005 \\
		model-seed2 & 0.095 & 0.115 & 0.203 \\
		model-seed3 & 0.093 & 0.111 & 0.199 \\
		model-seed4 & 0.094 & 0.108 & 0.200 \\
		Average & \textbf{0.101} & \textbf{0.112} & \textbf{0.152} \\
		\hline
	\end{tabular}
	\label{tab:efs_average}
\end{table}

\begin{figure}[H]
	\centering
	\includegraphics[width=.8\textwidth]{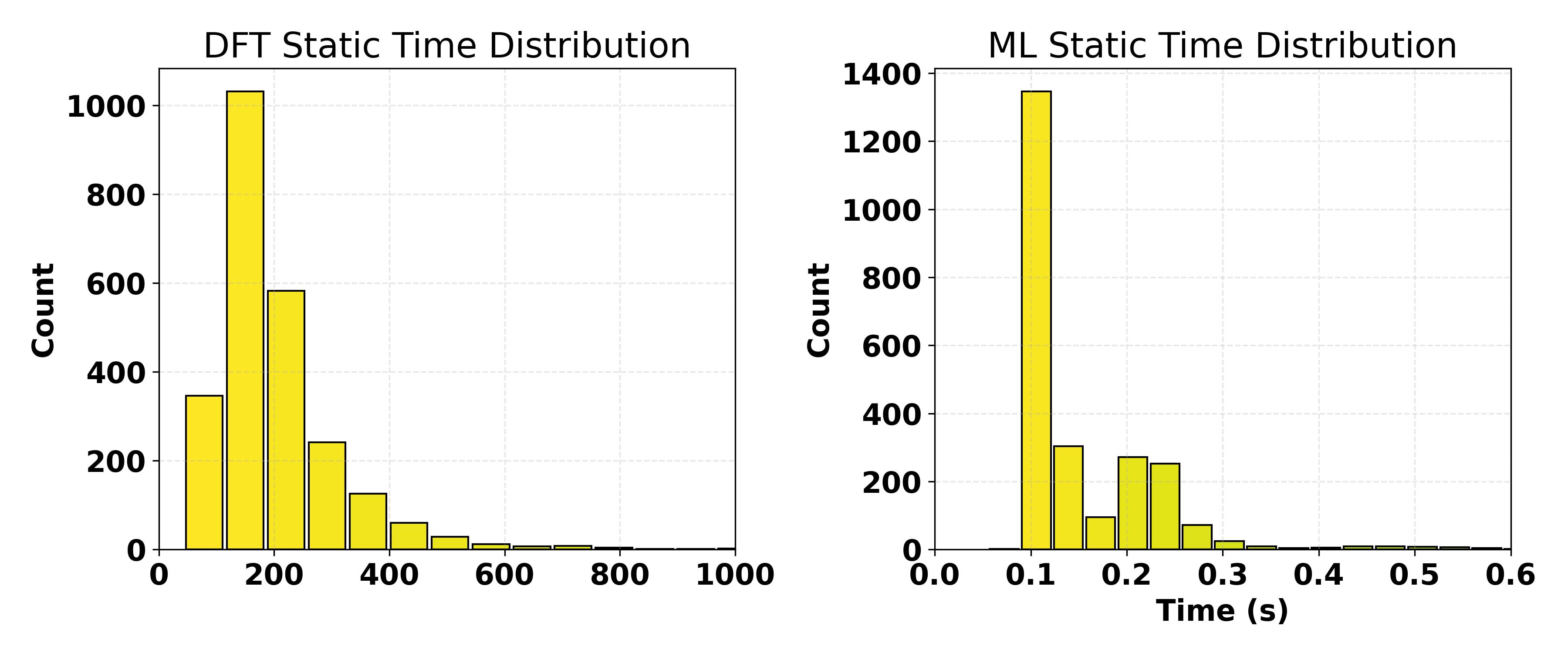}
	\caption{
		Comparison of computational time distributions for single-point static calculations using DFT and the ML model. The left panel shows the distribution of wall-clock time for DFT calculations across all structures, while the right panel displays the corresponding ML evaluation times.
	}
	\label{fig:dft_vs_ml_time}
\end{figure}

Figure~\ref{fig:dft_vs_ml_time} compares the computational time distributions for single-point static calculations using DFT and the ML model. Both DFT and ML calculations were performed on the same CPU node equipped with Intel Xeon Gold 6248R processors (48 cores, 2.60 GHz) to ensure a fair comparison.

\newpage
	\section{Properties simulation}
	To further evaluate the reliability of elastic property predictions, we performed an outlier filtering analysis to remove data points with large deviations. The original dataset contains 7,297 samples, of which 375 (5.1\%) and 401 (5.5\%) outliers were removed for the effective longitudinal modulus ($L_m$) and shear modulus ($G_m$), respectively. After filtering, the correlation coefficients for $L_m$ improved from 0.533 (identity) and 0.734 (linear) to 0.638 and 0.842, respectively. Similarly, the linear correlation coefficient for $G_m$ increased from 0.580 to 0.780. These results indicate that removing a small fraction of outliers significantly improves prediction consistency, suggesting that the majority of predictions are in reasonable agreement with DFT references, while a limited number of outliers strongly affect the overall statistics. This behavior is consistent with the fact that elastic constants, as second derivatives of total energy with respect to strain, are particularly sensitive to error propagation.
		\begin{figure}[htbp]
		\centering
		\includegraphics[width=0.95\textwidth]{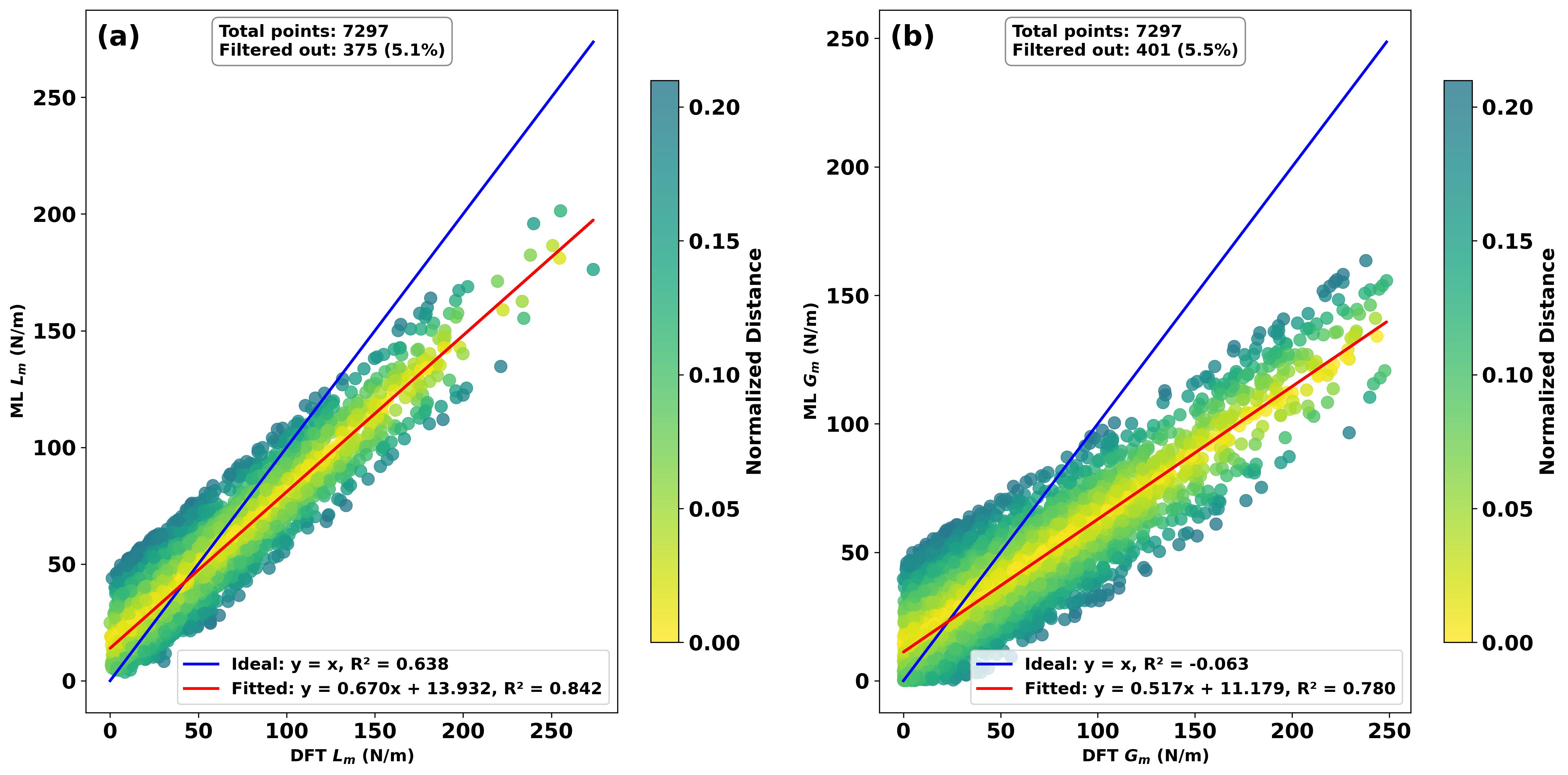}
		\caption{
			Elastic modulus comparison between DFT and ML predictions after outlier filtering. 
			(a) Longitudinal modulus $L_m$ and (b) shear modulus $G_m$. Points are colored by 
			normalized distance to fitted line. Filter statistics and $R^2$ values are shown in each subplot.    }
		\label{fig:elastic_modulus}
	\end{figure}
	From the filtered scatter distributions, both $L_m$ and $G_m$ predictions exhibit good linear relationships with the corresponding DFT values, indicating that the model captures the relative trends in elastic properties. However, the fitted lines still deviate from the ideal $y = x$ relationship, suggesting the presence of systematic bias. This deviation may originate from differences in exchange–correlation treatments and vDW interactions used in the reference data. In particular, the C2DB database calculations do not include vDW corrections, and the magnitude of vDW interactions varies across different material classes, which can lead to systematic deviations from the ideal linear relationship.

The equation of state (EOS) describes the fundamental relationship between the total energy of a material system and its volume or strain, and serves as an important method for evaluating the thermodynamic stability and mechanical properties of crystals. In this study, we focus on two-dimensional materials, where EOS curves are obtained by systematically applying biaxial strain to their equilibrium structures and calculating the corresponding total energies using both DFT and ML-IAP.

		\begin{figure}[H]
			\centering
			\includegraphics[width=0.9\textwidth]{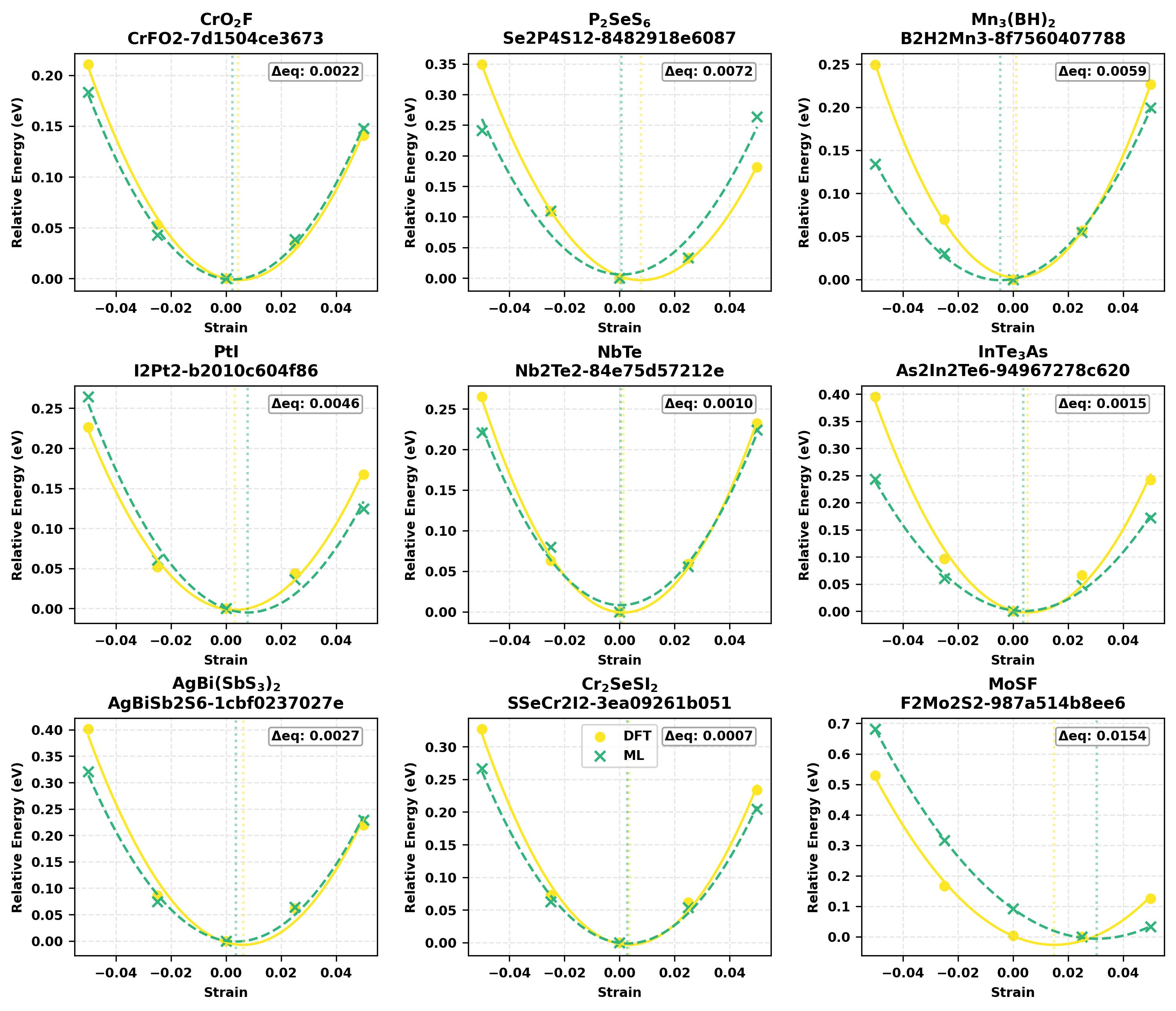}
			\caption{
				EOS comparison between DFT and machine learning model predictions for 9 randomly selected structures. The DFT energies (solid lines and circles) and ML energies (dashed lines and crosses) are plotted as functions of strain. Relative energies
				are referenced to the minimum energy of each method. Vertical dashed lines indicate the predicted equilibrium strain positions. The values $\Delta_\mathrm{eq}$ quantify the difference in equilibrium strain between DFT and ML for each structure.
			}
			\label{fig:eos_comparison}
		\end{figure}
	
In our previous work\cite{wang2025M4X8}, we performed high-throughput density functional theory (DFT) calculations to systematically investigate 84 V-shaped corrugated \ce{M4X8} monolayers (where M is a transition metal and X is a halogen), identifying 17 candidates with outstanding thermodynamic, dynamic, thermal, and mechanical stability. Notably, the DFT- and ML-predicted average phonon frequencies for these stable \ce{M4X8} compounds exhibit a strong linear correlation, underscoring the reliability of ML models in capturing vibrational properties.
	
		\begin{figure}[H]
			\centering
			\includegraphics[width=0.6\textwidth]{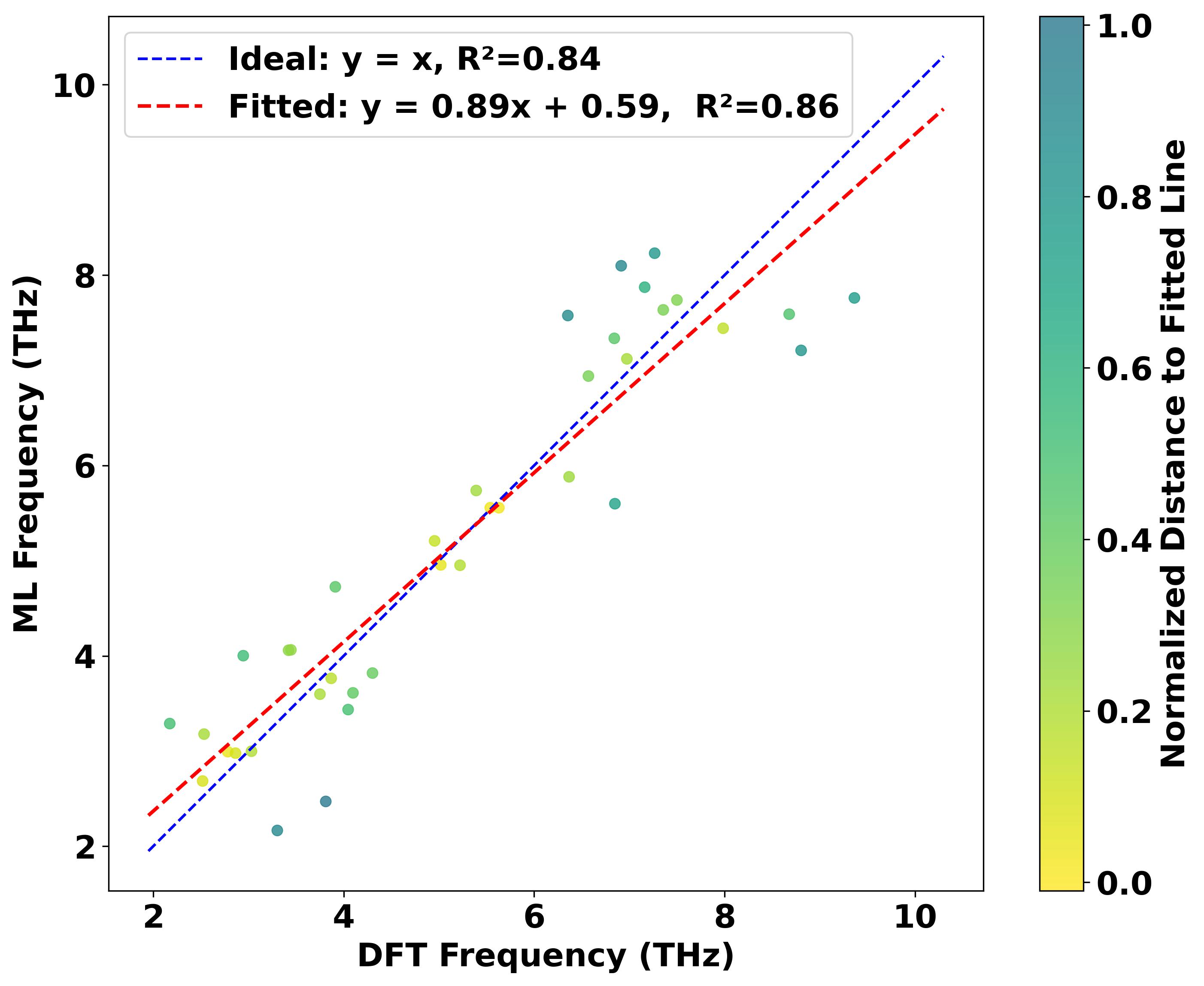}
			\caption{
				Comparison between DFT and ML-predicted average phonon frequencies for \ce{M4X8}  structures.  Each point represents a material, colored by the normalized perpendicular distance to the fitted regression line ($y = kx + b$).  The dashed blue line indicates the ideal relation $y = x$, while the dashed red line shows the best linear fit. }
			\label{fig:M4X8_dft_ml_comparison}
		\end{figure}
	
\begin{figure}[htbp]
	\centering
	\includegraphics[width=0.6\textwidth]{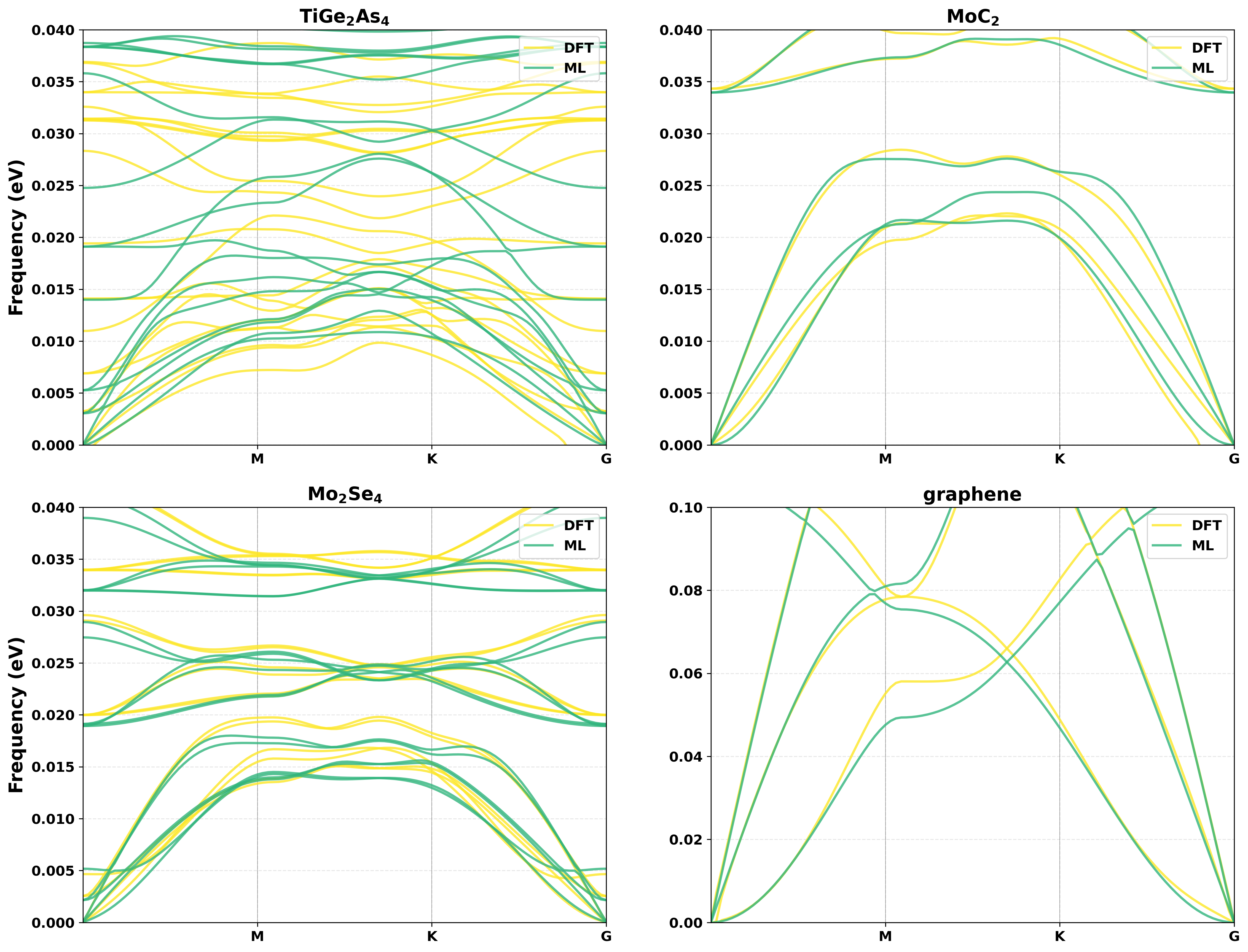}
	\caption{Phonon band structure comparison between DFT and ML predictions for four representative 2D materials: (a) TiGe$_2$As$_4$, (b) MoC$_2$, (c) Mo$_2$Se$_4$, and (d) graphene.}
	\label{fig:phonon_comparison}
\end{figure}

\newpage
\section{Molecular dynamics simulation}

The diffusion behavior of lithium ions within the interlayer nanochannels of bilayer \ce{MoS2} was characterized using the mean square displacement (MSD) obtained from equilibrium NVT ensemble simulations. The diffusion coefficient $D$ was calculated based on the in-plane MSD along the \ce{MoS2} basal planes, following the relation:
\begin{equation}
\langle x(t)^2 \rangle + \langle y(t)^2 \rangle = 4Dt
\end{equation}
where $\langle x(t)^2 \rangle$ and $\langle y(t)^2 \rangle$ are the average squared displacements along the $x$ and $y$ directions, respectively.
According to the Arrhenius equation, the temperature dependence of the diffusion coefficient $D$ is expressed as:
\begin{equation}
D(T) = D_0 \exp\left(-\frac{E_a}{k_B T}\right)
\end{equation}
where $D_0$ is the pre-exponential factor (also known as the frequency factor), $E_a$ is the activation energy, and $k_B$ is the Boltzmann constant.
The diffusion of ions within the nanochannels exhibits a linear relationship between $\ln D$ and $1000/T$, from which the activation energy $E_a$ can be extracted from the slope.

\begin{figure}[htbp]
	\centering
	\includegraphics[width=0.82\textwidth]{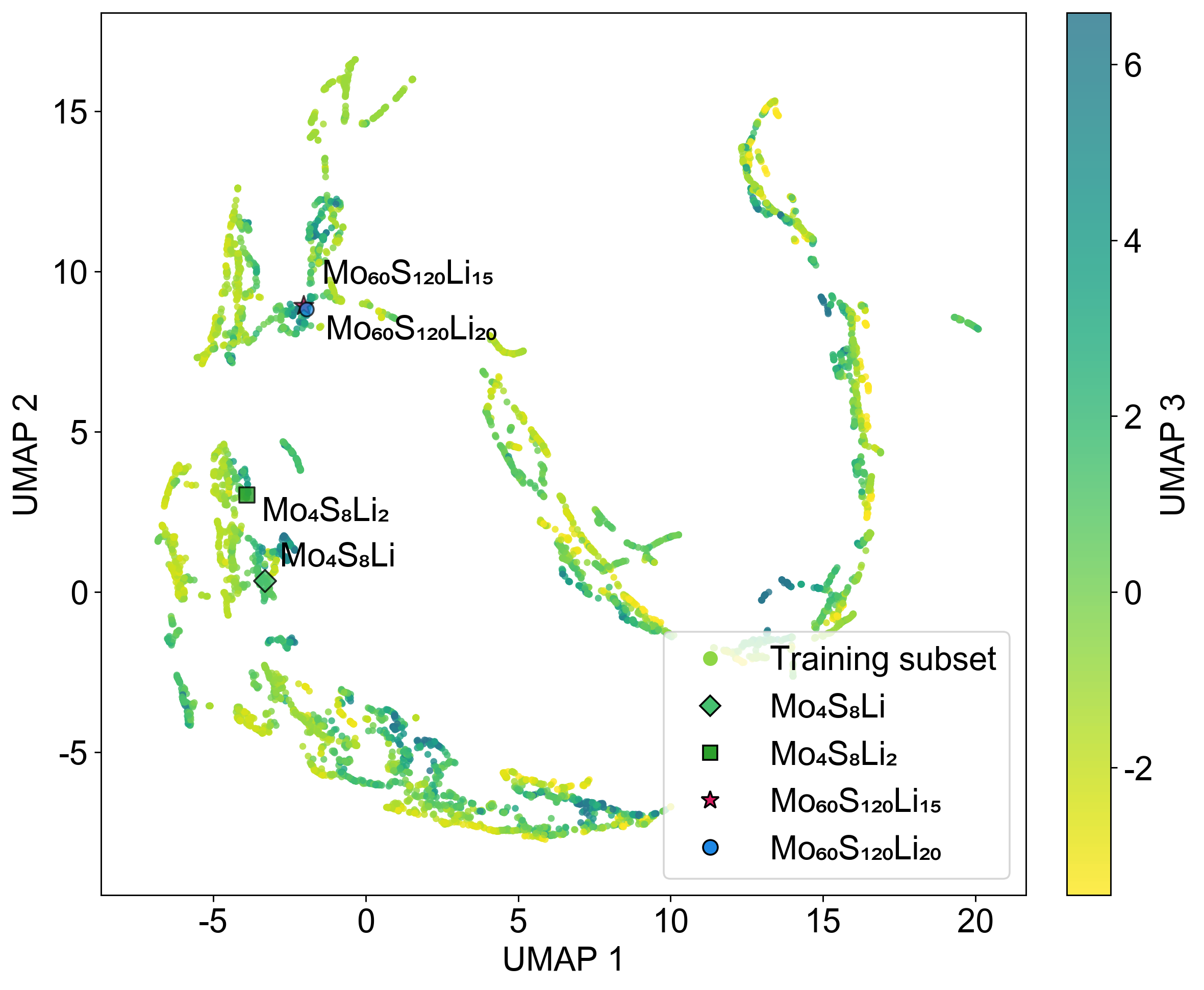}
	\caption{UMAP embedding of the training structures together with representative Li-intercalated reference structures, Mo$_4$S$_8$Li and Mo$_4$S$_8$Li$_2$, and the two MD-derived structures, Mo$_{60}$S$_{120}$Li$_{15}$ and Mo$_{60}$S$_{120}$Li$_{20}$. The horizontal axis (UMAP 1) and vertical axis (UMAP 2) represent the first two coordinates of the low-dimensional embedding of the structure descriptor space, while the color scale (UMAP 3) represents the third embedding coordinate.}
	\label{fig:umap_md_training}
\end{figure}

To evaluate the relationship between the MoS$_2$--Li systems and the sub-training data, we performed an analysis based on SOAP \cite{Bartok2013}
descriptors combined with dimensionality reduction, as shown in Figure~\ref{fig:umap_md_training}. The analysis reveals that the Li--Mo--S training dataset exhibits several distinct clusters in descriptor space. The MoS$_2$Li and Mo$_4$S$_8$Li$_2$ structures lie within the distribution of the training data, whereas Mo$_{60}$S$_{120}$Li$_{15}$ and Mo$_{60}$S$_{120}$Li$_{20}$ are located near the boundary of the training distribution. This suggests that the MD systems studied here differ from the training data, particularly for high lithium concentration configurations, which are sparsely represented in the training dataset.

\begin{table}[htbp]
	\centering
	\caption{Descriptor-space distance and local density metrics relative to the training subset.}
	\label{tab:mo-s-li}
	\begin{tabular}{lcc}
		\hline
		Structure & 5-NN mean distance & Local density percentile (\%) \\
		\hline
		\ce{MoS2Li}        & 1.648  & 42.66 \\
		\ce{Mo4S8Li2} & 1.765 & 30.95 \\
		\ce{Mo60S120Li15}   & 12.248 & 0.44 \\
		\ce{Mo60S120Li20}   & 12.329 & 0.43 \\
		\hline
	\end{tabular}
\end{table}
To further quantify this difference, we computed descriptor-space distance and local density metrics (see Table~\ref{tab:mo-s-li}). The 5-NN mean distance represents the average distance to the five nearest training structures, while the local density percentile indicates the relative density of the surrounding region within the training dataset. The results show that both MoS$_2$Li and Mo$_4$S$_8$Li$_2$ exhibit relatively small 5-NN mean distances (1.648 and 1.765, respectively) and moderate density percentiles (42.66\% and 30.95\%), indicating that these structures lie within well-sampled regions of the training space. In contrast, Mo$_{60}$S$_{120}$Li$_{15}$ and Mo$_{60}$S$_{120}$Li$_{20}$ exhibit significantly larger 5-NN distances (12.248 and 12.329, respectively) and extremely low density percentiles (0.44\% and 0.43\%), suggesting that these structures lie in sparsely sampled regions of the training distribution and do not directly overlap with the training data.

\newpage
\section{Highthrough-put screening}
        
To perform high-throughput screening of the $\alpha_1$-\ce{MA2Z4} structures, we use four base models with different random initializations to construct an ensemble model. During the screening process, the structures were first optimized to filter out those that did not satisfy symmetry constraints. Elastic property calculations were then performed based on the Born stability criteria to identify elastically stable structures. Finally, phonon calculations were conducted to determine dynamical stability by checking for the absence of imaginary frequencies. Through this three-step screening procedure, a total of 539 stable structures were obtained. Subsequently, density functional theory (DFT) calculations of the energy above the convex hull were carried out, resulting in 12 structures that remained stable at this level.

\begin{table}[H]
\centering
\caption{Statistics of stability-based structure filtering using model ensembles, including elastic, phonon, and DFT DFT E$_{hull}$ criteria.}
\label{tab:model_compare}
\begin{tabular}{lcccc}
\toprule
\textbf{Properties} & \textbf{model-1} & \textbf{model-2} & \textbf{model-3} & \textbf{model-4} \\
\midrule
Relaxation   & 1676 & 1685 & 1684 & 1682 \\
Elastic      & 1667 & 1680 & 1673 & 1673 \\
Phonon       & 1045 & 973 & 978  & 859  \\
Intersection & \multicolumn{4}{c}{536} \\
DFT E$_{hull}$           & \multicolumn{4}{c}{12} \\
\bottomrule
\end{tabular}
\end{table}

    \begin{figure}[H]
    \centering
    \includegraphics[width=0.8\textwidth]{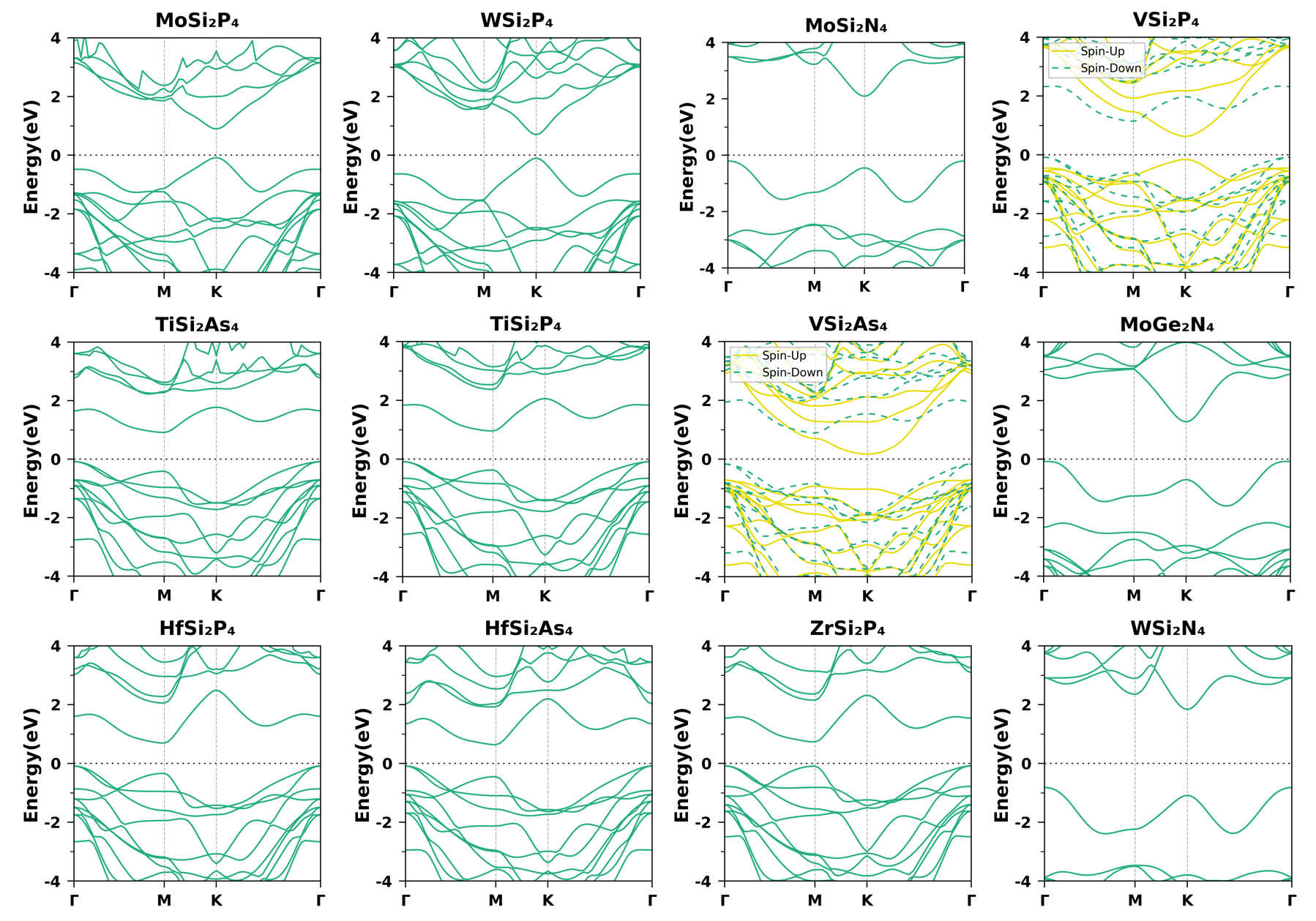}
    \caption{Calculated electronic band structures (DFT, GGA+U) for stable compound.
    The Fermi level is set to 0 eV and is indicated by the dashed horizontal line. Vertical dashed lines indicate high-symmetry points in the Brillouin zone.}
    \label{fig:ma2z4_band_structures}
\end{figure}
		\begin{figure}[H]
    \centering
    \includegraphics[width=0.8\textwidth]{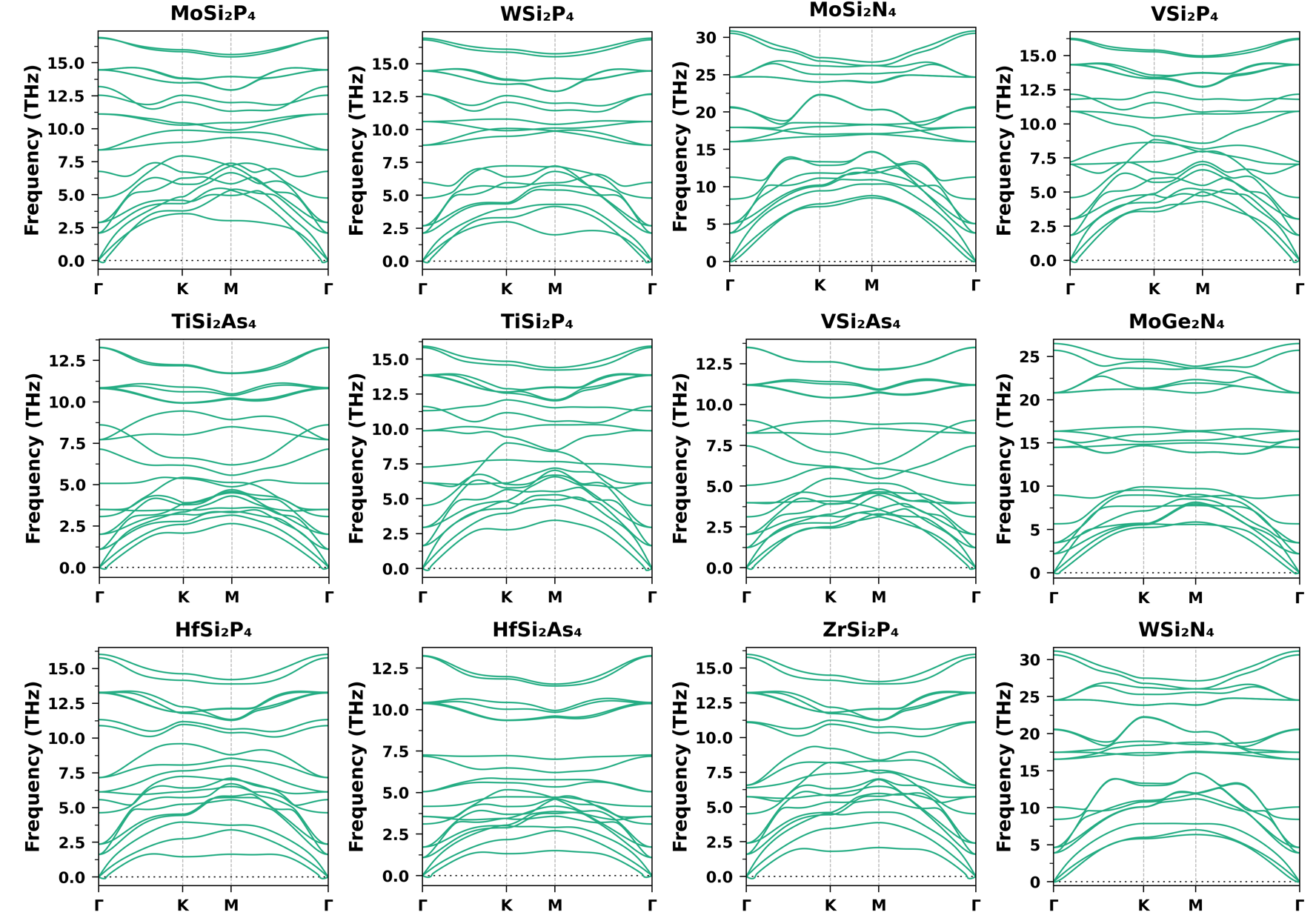}
    \caption{Calculated phonon band structures (DFT, GGA+U) for 12 stable \ce{MA2Z4} compounds. Vertical dashed lines indicate high-symmetry points in the Brillouin zone.}
    \label{fig:ma2z4_phonon_bands}
\end{figure}

\newpage
\section{Material simulation agent}

The Material Simulation Agent (MatAgent) is an automated service designed for materials simulations, built on the UniMatSim framework and integrated with the Uni2D machine learning interatomic potential developed in this work. Given a material ID, the system automatically resolves the corresponding structure file, loads the specified potential model, and performs simulation tasks such as structural optimization, phonon calculations, elastic property evaluation, molecular dynamics simulations, band gap, and equation-of-state analysis. 

All simulation results are provided in a structured format, including JSON data files, visualization figures, and automatically generated analysis reports. The computational parameters and runtime environment metadata are also recorded to ensure traceability and reproducibility of the simulation workflow.

\begin{figure}[htbp]
	\centering
	\includegraphics[width=1\textwidth]{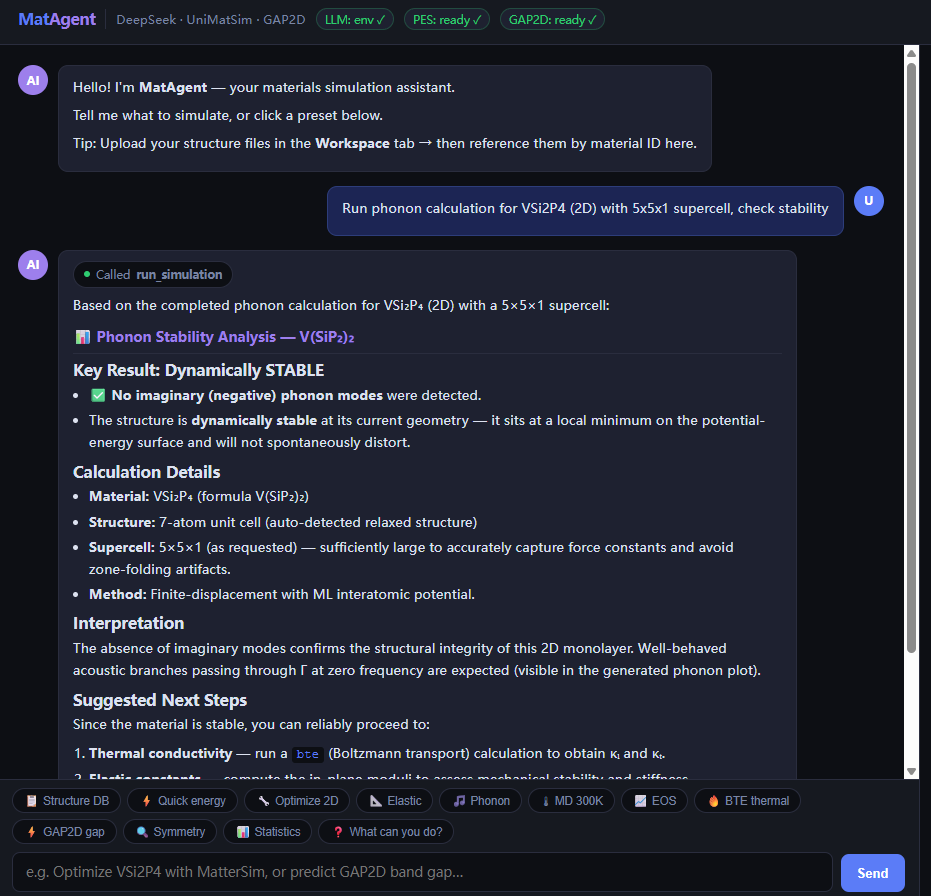}
	\caption{User interface of the agent showing the interactive dialogue window.}
	\label{fig:mcp-agent-1a}
\end{figure}

\clearpage

\begin{figure}[htbp]
	\centering
	\includegraphics[width=1\textwidth]{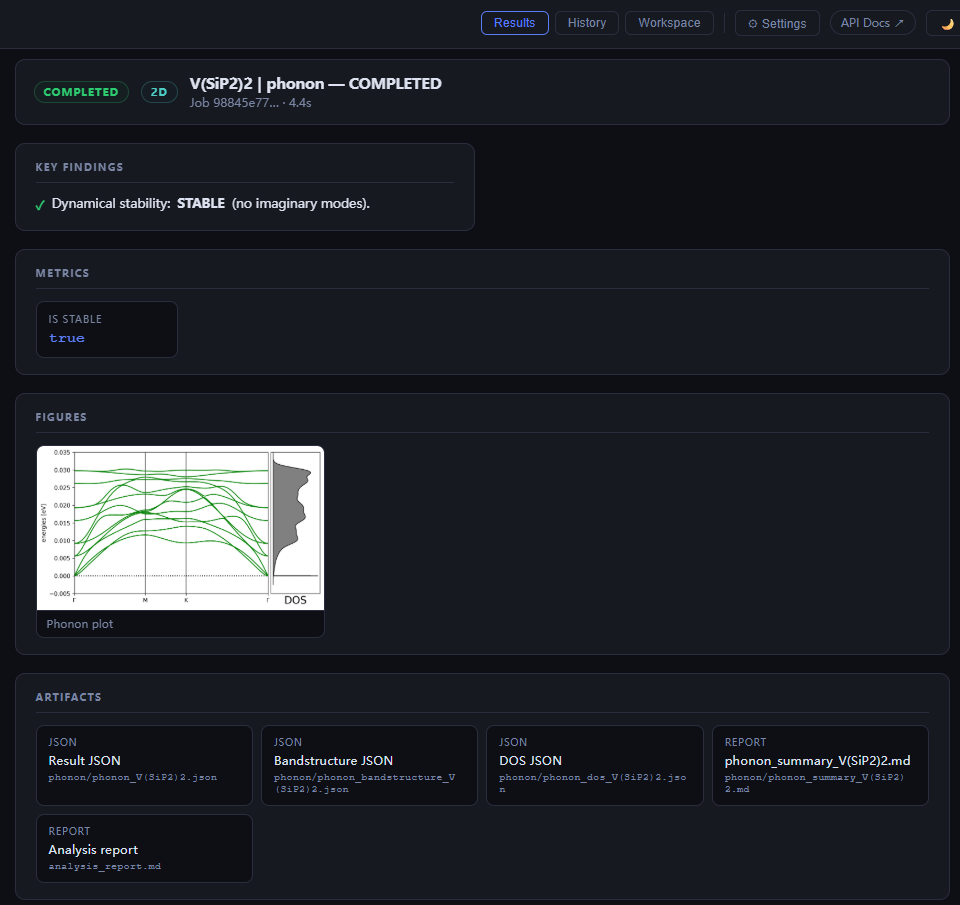}
	\caption{Simulation results visualization interface generated by the agent.}
	\label{fig:mcp-agent-1b}
\end{figure}

\newpage
\section{Electronic Band Gap Model}

The Gap2D model is developed based on a graph neural network (GNN) architecture for predicting the electronic band gaps of two-dimensional materials. In this framework, classification and regression tasks are treated as two independent models that are trained and evaluated separately. Crystal structures are first converted into graph representations, where atoms are treated as nodes and neighboring atomic interactions are represented as edges. Node features are constructed using atomic number embeddings, while edge features are generated through a combination of spherical Bessel radial functions and spherical harmonics, enabling the model to encode both local geometric environments and directional information. The graph representations are processed using multiple message-passing layers, followed by global mean and max pooling to obtain structure-level features. The classification model predicts whether a material possesses a finite band gap, whereas the regression model independently predicts the band gap value.

The performance of the Gap2D model was evaluated using 20-fold cross-validation for both band-gap regression and metal–semiconductor classification tasks with C2DB dataset. For the regression task, the model achieves an average MAE of 0.532~eV with a standard deviation of 0.117~eV, while the best-performing fold reaches an MAE of 0.403~eV. For the classification task, the model attains an average accuracy of 0.849 and an F1 score of 0.853, with the best-performing fold achieving an accuracy of 0.880. These results indicate stable predictive performance across different data splits. The figures presented below correspond to the best-performing models.

\begin{figure}[H]
	\centering
	\includegraphics[width=\textwidth]{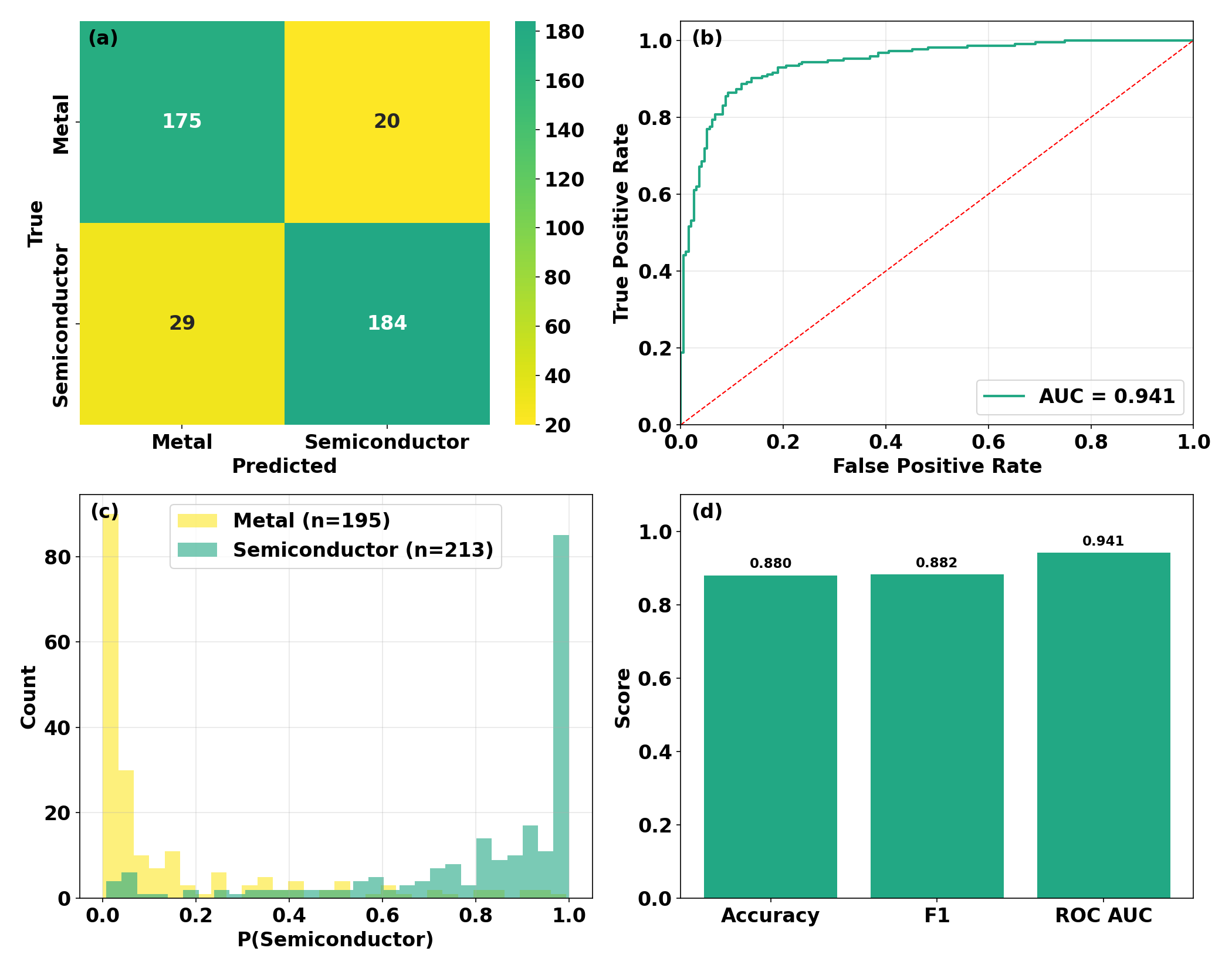}
	\caption{
		Evaluation results of the GNN-based classification model on the metal/semiconductor dataset. 
		(a) Confusion matrix showing predicted versus true labels for metals and semiconductors. 
		(b) Receiver operating characteristic (ROC) curve with the corresponding AUC value. 
		(c) Distribution of predicted semiconductor probabilities separated by true class labels. 
		(d) Summary of classification metrics including accuracy, F1 score, and ROC-AUC. 
		A threshold of 0.005~eV was used to distinguish metals and semiconductors.
	}
	\label{fig:classification-metrics}
\end{figure}

The classification performance of the best model is summarized in Figure~\ref{fig:classification-metrics}. The confusion matrix shows that most samples are correctly classified, with 175 metals and 184 semiconductors correctly identified. The model achieves an accuracy of 0.880, an F1 score of 0.882, and an ROC-AUC value of 0.941, indicating strong discrimination capability between metallic and semiconducting materials. The probability distribution further demonstrates clear separation between the two classes, suggesting reliable classification confidence.

\begin{figure}[H]
	\centering
	\includegraphics[width=\textwidth]{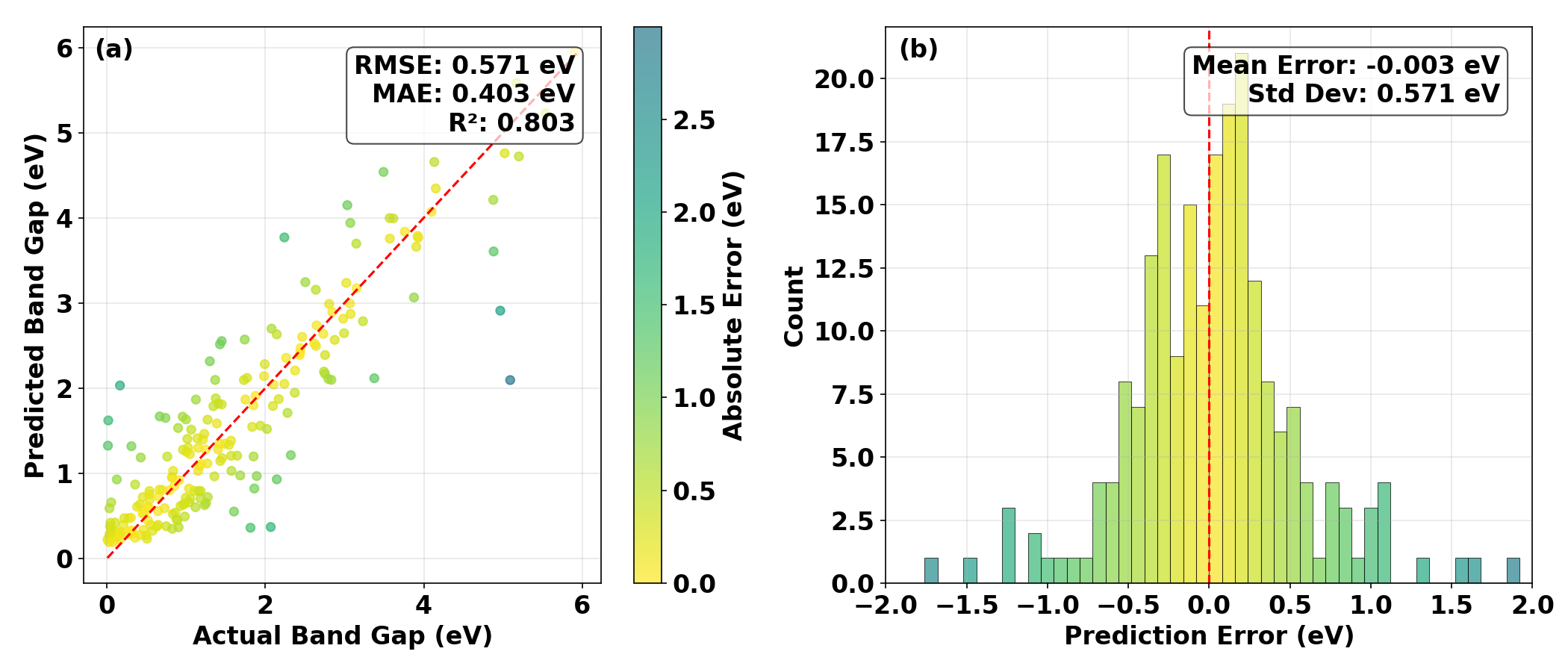}
	\caption{
		Regression performance of the GNN-based model for band-gap prediction of 2D materials. 
		(a) Scatter plot of predicted versus reference band-gap values, where the color represents the absolute prediction error. 
		(b) Distribution of prediction errors, including mean error and standard deviation.
	}
	\label{fig:regression-metrics}
\end{figure}

The regression performance of the best model for band gap prediction is presented in Figure~\ref{fig:regression-metrics}. The model achieves an MAE of 0.403~eV and an RMSE of 0.571~eV, with a coefficient of determination $R^2$ of 0.803. The error distribution is centered around zero with a mean error of $-0.003$~eV and a standard deviation of 0.571~eV, indicating unbiased predictions and stable performance.

	\footnotesize{
		\bibliography{main}
	}